\documentclass{article}

\usepackage{arxiv}

\usepackage[utf8]{inputenc} 
\usepackage[T1]{fontenc}    
\usepackage{hyperref}       
\usepackage{url}            
\usepackage{booktabs}       
\usepackage{amsfonts}       
\usepackage{nicefrac}       
\usepackage{microtype}      
\usepackage{lipsum}
\usepackage{graphicx}
\usepackage{multirow}
\usepackage{bm}
\usepackage{mathptmx}
\usepackage{dcolumn}
\usepackage{etoolbox}
\usepackage[dvipsnames]{xcolor}
\usepackage{amsmath,bm}

\graphicspath{ {./images/} }

\title{Modeling Wind Turbine Performance and Wake Interactions with Machine Learning}

\author{
 Coleman Moss \\
  Wind Fluids and Experiments (WindFluX) Laboratory\\
  The University of Texas at Dallas\\
  Richardson, TX 75080 \\
  \texttt{coleman.moss@utdallas.edu} \\
   \And
 Giacomo Valerio Iungo \\
  Wind Fluids and Experiments (WindFluX) Laboratory\\
  The University of Texas at Dallas\\
  Richardson, TX 75080 \\
  \texttt{valerio.iungo@utdallas.edu} \\
   \And
 Romit Maulik \\
  Argonne National Laboratory\\
  Lemont, IL 60439 \\
}

\begin{document}
\maketitle
\begin{abstract}
Different machine learning (ML) models are trained on SCADA and meteorological data collected at an onshore wind farm and then assessed in terms of fidelity and accuracy for predictions of wind speed, turbulence intensity, and power capture at the turbine and wind farm levels for different wind and atmospheric conditions. ML methods for data quality control and pre-processing are applied to the data set under investigation and found to outperform standard statistical methods. A hybrid model, comprised of a linear interpolation model, Gaussian process, deep neural network (DNN), and support vector machine, paired with a DNN filter, is found to achieve high accuracy for modeling wind turbine power capture. Modifications of the incoming freestream wind speed and turbulence intensity, $TI$, due to the evolution of the wind field over the wind farm and effects associated with operating turbines are also captured using DNN models. Thus, turbine-level modeling is achieved using models for predicting power capture while farm-level modeling is achieved by combining models predicting wind speed and $TI$ at each turbine location from freestream conditions with models predicting power capture. Combining these models provides results consistent with expected power capture performance and holds promise for future endeavors in wind farm modeling and diagnostics. Though training ML models is computationally expensive, using the trained models to simulate the entire wind farm takes only a few seconds on a typical modern laptop computer, and the total computational cost is still lower than other available mid-fidelity simulation approaches.
\end{abstract}


\section{Introduction}\label{sec:introduction}
The wind energy sector continues to grow rapidly to meet the ever-growing global energy need and to address the rising concerns associated with fossil fuel use, such as eventual depletion and current greenhouse gas emissions. High targets are set for future global capacity \cite{Kaldellis2011} while the current cost of production drops \cite{Gielen2019}. To continue adding to the global wind energy capacity in an efficient manner, great attention has been given to the design and control of wind turbines operating individually in a ``greedy'' manner, and when operating collectively in a wind farm. These tasks imply the use of models to predict wind turbine operations with great accuracy yet also demand that the related computational costs be kept small, considering the large number of wind farm simulations required for the design and optimization of a single wind power plant \cite{Gonzalez2010,Santhanagopalan2018,Gu2021}.

To minimize the levelized cost of energy (LCOE) of a wind farm, it is necessary to quantify turbine performance in relation to environmental conditions, such as wind speed at the turbine hub height. A wind turbine power curve (WTPC hereinafter) quantifies this relationship and has a standardized approach for its derivation \cite{IEC2017}. In addition, several other approaches have been proposed as alternatives to achieve more accurate power predictions \cite{Wang2019}. While turbine manufacturers typically provide a WTPC, every turbine performs differently in unique topographic and environmental conditions, so the best approach to quantifying the performance of a given turbine is deemed to be utilizing operational data coupled with local wind measurements \cite{Lydia2013,Zhou2014,Wagner2014}. The supervisory control and data acquisition (SCADA) system is a good source of this data, though it often contains outliers due to, among other causes, turbine shutdown for maintenance, turbine de-rating, icing, or sensor fault. These outliers have a high likelihood of negatively impacting the accuracy of models trained using the data, thus the SCADA data needs to be filtered \cite{Qiao2021,Sainz2009}. Once accurate models are developed, the energy production for a wind farm can be estimated \cite{Schallenberg2013}. Additionally, good WTPC models can reduce wind farm costs by providing effective means of monitoring turbine operating conditions and detecting off-design behaviors \cite{Marvuglia2012}.

Wind turbine modeling typically encompasses three steps: processing the available experimental data of turbine power capture, modeling the WTPC, and applying the model for power performance predictions. Many approaches are available for each of these steps. Manual approaches to processing data have been proposed \cite{Meik2011}, yet these are typically not favored. Automatic filtering approaches using statistical methods to reject outliers are more common. For instance, a simple approach could be to bin data \cite{IEC2017} and reject points further than a prescribed multiple of the data standard deviation associated with the respective bin. A more sophisticated approach is to use K-means clustering and the Mahalanobis distance to cluster data points and reject outliers \cite{Yesilbudak2018}. The K-means clustering can also be replaced with Fuzzy C-Means clustering \cite{Pei2019}. A benefit of filters utilizing the Mahalanobis distance is that it measures distances between data points along multiple dimensions. Other statistical approaches that have been proposed include a least median squares approach \cite{Sainz2009} and a standard deviation approach using exponential smoothing to make the wind speed time series stationary \cite{Ouyang2017}. A drawback of statistical filters is that they impose the \textit{a priori} assumption that good data are close to a cluster centroid. By using a machine learning (ML hereinafter) approach to filtering, this assumption is replaced with the assumption that the ML model accurately captures turbine performance such that good data points are predictable with the model. One approach in this direction is to use a Gaussian process (GP hereinafter) model, which natively reports a confidence interval associated with every prediction, so outliers can readily be rejected if they fall outside the confidence interval \cite{Manobel2018}. Autoencoder networks, which find low-dimensional representations of high-dimensional snapshots of data, were used to filter SCADA data based on assumed groupings in the reduced-dimensional space \cite{Neshat2021}.

Approaches to model WTPCs can be broken down into two categories: parametric and non-parametric. Parametric WTPC models generally fit equations estimating the WTPC to given data by tuning different parameters in the equation. The simplest parametric approach breaks the WTPC into several smaller linear segments, giving a piecewise definition of the WTPC that is tuned by adjusting each smaller line segment \cite{Wang2019}. Other parametric approaches take advantage of the cubic relationship between wind speed and power \cite{Deshmukh2008} or Weibull's parameters \cite{Powell1981}. Sigmoid and Gaussian cumulative distribution function approaches have also been used to specifically model region two of the WTPC, namely between cut-in wind speed and rated wind speed of a wind turbine \cite{Osadciw2010}. New developments in parametric modeling typically involve new functions to be fitted from the SCADA data, such as the modified hyperbolic tangent \cite{Taslimi2016}. Non-parametric modeling approaches encompass a data-driven ML approach. Common ML regression models include support vector machines \cite{Ouyang2017} (SVM) and GPs \cite{Zhou2014}. Deep Neural Networks (DNN) have been shown to accurately predict wind turbine performance using SCADA data \cite{Wu2021}. Another work \cite{Veena2020} showed that SVM models performed better than Artificial Neural Network (ANN), K-Nearest Neighbor (KNN), and Multivariate Adaptive Regression Spline (MARS) models for SCADA data for a specific site, though all these models performed better than the manufacturer-provided WTPC, and that the SVM model accurately predicted the performance of a different turbines in a different site, since ``the features of the site and corresponding flow conditions at different wind velocities are embedded in such data-driven models.'' Data-driven models can also capture power variations due to turbulence effects \cite{Clifton2014}, which are known to impact power production \cite{StMartin2016, Bardal2017}, as well as air density fluctuations \cite{Pandit2019}.

WTPC models can be applied in a variety of ways. A simple method is to use the model to predict power production for a validation set of environmental conditions and compare the predicted and actual power. The annual energy production (AEP) can be calculated according to the IEC standard \cite{IEC2017} for a given power curve, and the estimated AEPs of different models can be compared with the IEC approach and the real AEP if power data is available. WTPC can also be probed to estimate turbine response to different environmental conditions. For instance, if a model is capable of accepting turbulence intensity ($TI$) or wind direction as inputs, different power curves can be produced by varying these parameters.

The contribution of this paper is to apply the above-mentioned approaches to filtering and modeling a SCADA data set from a single wind farm. Specifically, the effectiveness of four different filters, i.e., a simple binning filter following the IEC standard \cite{IEC2017}, a K-means clustering with Mahalanobis distance filter \cite{Yesilbudak2018}, a GP filter \cite{Manobel2018}, and a novel DNN filter, is investigated. The filters will be applied to pre-processed SCADA data to remove outliers. Two approaches using four different models are then used to model phenomena of interest. With the first approach, the turbine performance is modeled using an ML model for the WTPC. This model considers unwaked performance across sets of turbines, which may be specific rows or the entire farm, in aggregate to provide a general power model and is referred to as the isolated-turbine approach. The second approach utilizes three different models to predict power capture at the turbine level by taking into account effects due to clustering multiple wind turbines in a wind farm, e.g. wake interactions, while using as inputs solely freestream conditions. Two models are used to predict the incoming wind speed and $TI$ at a given turbine as a function of reference conditions, which approximate freestream conditions. These models capture turbine-to-turbine interactions, such as wakes and speedups, as well as environmental effects. Finally, individual turbine power models predict power capture from the predicted wind speed and $TI$.

The remainder of the paper is organized as follows. In section \ref{sec:data set_farm_overview}, the farm under consideration is introduced along with the SCADA data set and pre-processing applied to the data. Section \ref{sec:filtering} explains the WTPC region removal applied to the input data set, as well as the four filters applied. All the models used for the isolated-turbine and clustered-turbine approaches are covered in section \ref{sec:modeling}. The performance of the models is validated qualitatively and verified quantitatively in sections \ref{sec:model_validation} and \ref{sec:model_verification}, respectively. Section \ref{sec:data_requirements} provides guidelines for the definition of metrics that can be used to determine if a given SCADA data set is suitable for WTPC ML modeling and to guide the design of field campaigns for creating data sets for ML modeling. Section \ref{sec:computational_cost} discusses cases where ML models outperform a mid-fidelity RANS solver in terms of computational cost. Lastly, section \ref{sec:conclusion} provides some final remarks.

\section{Data Set and Wind Farm Overview}\label{sec:data set_farm_overview}
\begin{figure*}[t!]
    \centering
    \includegraphics[width=\textwidth]{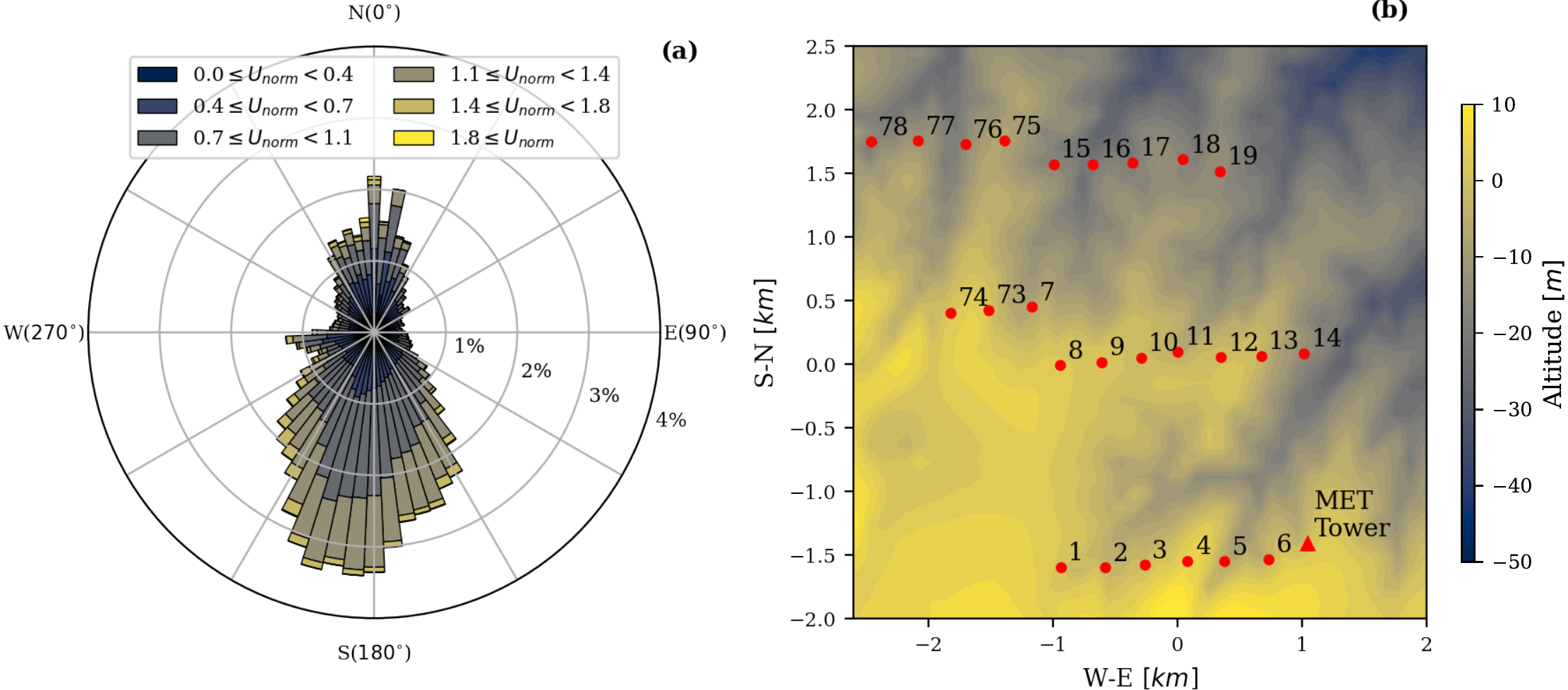}
    \caption{Characterization of the Panhandle wind farm: (a) Windrose of the site with wind speed normalized by the turbine rated wind speed of 11 m$\cdot$s$^{-1}$; (b) layout of the wind farm.}
    \label{fig:layout_and_windrose}
\end{figure*}

The wind farm under consideration is located in the Panhandle of Texas and includes 25 wind turbines arranged in three rows roughly aligned along the East-West direction \cite{El-Asha2017,Zhan2020,Zhan2020WES,Iungo2022}. The details of the turbines are summarized in Table \ref{tab:turbine_details} while the site wind rose and wind farm layout are reported in figures \ref{fig:layout_and_windrose}(a) and \ref{fig:layout_and_windrose}(b), respectively.

\begin{table*}[b]
    \centering
    \begin{tabular}{|c|c|c|c|}
        \hline
        \textbf{Manufacturer} & Siemens & \textbf{Model} & SWT-2.3-108 \\ \hline
        \textbf{Year Online} & 2014 & \textbf{Rated Capacity} & 2.3 MW \\ \hline
        \textbf{Hub Height} & 80 m & \textbf{Rotor Diameter} & 108 m \\ \hline
        \textbf{Cut In Wind Speed} & 3 m$\cdot$s$^{-1}$ & \textbf{Rated Wind Speed} & 11 m$\cdot$s$^{-1}$\\ \hline
        \textbf{Cut Out Wind Speed} & 25 m$\cdot$s$^{-1}$ & \textbf{No. Turbines} & 25 \\ \hline
    \end{tabular}
    \caption{Technical details of the wind turbines at the Panhandle wind farm.}
    \label{tab:turbine_details}
\end{table*}

Meteorological data were collected from a meteorological (met)-tower starting on July 17\textsuperscript{th}, 2014, and continuing until June 23\textsuperscript{rd}, 2017. Specifically, time stamp, wind speed, wind direction, ambient temperature, pressure, and air density are provided as mean and standard deviation over 10-minute periods. After not-a-number (\textit{NaN}) data rejection, a total of 150,132 time stamps are available, which correspond to a total time of 2.85 years.

SCADA data for every turbine were recorded from August 20\textsuperscript{th}, 2015 up until April 15\textsuperscript{th}, 2017. The data is composed of statistics over 10-minute periods for wind speed, $TI$ (defined as the ratio between the wind speed standard deviation and its mean value), wind direction, ambient temperature, and power capture. The number of \textit{NaN} data points varies from turbine to turbine. After removing these points, the turbine with the highest number of down-selected samples has 65,034 points (1.24 years of data). The turbine with the least amount of data has 59,566 down-selected samples (1.13 years of data). The average number of down-selected samples per turbine is 63,517 giving 1.21 years of data. These data are not necessarily continuous in time, though, as removing time stamps with \textit{NaN} values causes discontinuities in time. Finally, the number of time stamps where all 25 wind turbines and the meteorological tower have non-\textit{NaN} values is 40,720 for 0.77 years of data.

\begin{figure*}
    \centering
    \includegraphics[width=\textwidth]{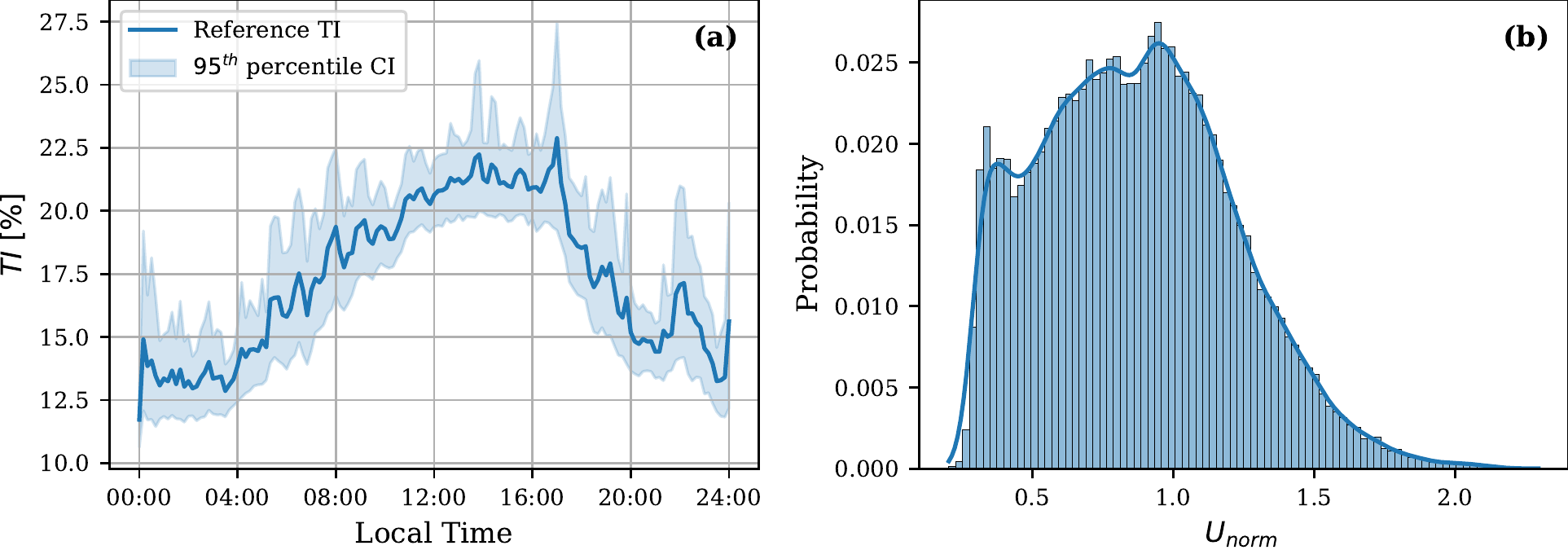}
    \caption{Climatology for the site of the Panhandle wind farm: (a) Daily cycle of $TI$ from the met-tower with \(95^{th}\) confidence interval calculated using bootstrapping; (b) Histogram of the reference hub-height wind speed.}
    \label{fig:ti_daily_cycle}
\end{figure*}

To investigate the climatology of this site, the daily cycle of $TI$ is reported in figure \ref{fig:ti_daily_cycle}(a). Since no instrument is available to provide actual freestream measurements, reference conditions are defined as the average of environmental conditions across all turbines not affected by wakes generated upstream (unwaked turbines), as proposed in a previous work studying the site under consideration\cite{El-Asha2017}. This procedure is used to define reference conditions for hub-height wind speed, wind direction, and $TI$. Reference $TI$ is found to have a strong daily cycle associated with the variation of atmospheric stability \cite{Magnusson1994,Hansen2012,Wharton2012,Abkar2015,Xie2017,Iungo2014}. The probability density function of the reference hub-height wind speed is reported in figure \ref{fig:ti_daily_cycle}(b). The SCADA wind speed is density-corrected as follows\cite{IEC2017}:
\begin{equation}\label{eqn:density_correction}
    U_{corr}=U \left( \frac{\rho}{\rho_0} \right)^{1/3},
\end{equation}
where \(\rho\) is the air density at each turbine, \(\rho_0\) is the reference air density of 1.225 kg$\cdot$ m$^{-3}$, and \(U\) is the mean wind speed. Since air density measurements are not available at each turbine, the best approximation is to calculate air density from the turbine ambient temperature in Kelvin, \(T\), and the met-tower pressure in Pascals, \(P\):
\begin{equation}\label{eqn:density_calculation}
    \rho=\frac{1}{T}\left[ \frac{P}{R_0} - \phi ~ a~e^{bT} \left( \frac{1}{R_0} - \frac{1}{R_w} \right) \right],
\end{equation}
where \(R_0\) is the gas constant of dry air (287.05 J $\cdot$ kg$^{-1} \cdot$ K$^{-1}$), \(R_w\) is the gas constant of water vapor (461.6 J $\cdot$ kg$^{-1} \cdot$ K$^{-1}$), \(\phi\) is the relative humidity (set to 0.5 since humidity measurements are unavailable), \(a\) is a constant equal to 0.0000205, and \(b\) is a constant equal to 0.0631846. A nacelle transfer function for the turbines is then calculated to correct turbine 06 since it is closest to the met-tower. This correction is applied to the other turbines through the same scale factor estimated for turbine 06 \cite{Letizia2022, Sebastiani2020}. In this way, bias errors in the turbine anemometers should be corrected or, at least, reduced.

\begin{figure*}
    \centering
    \includegraphics[width=\textwidth]{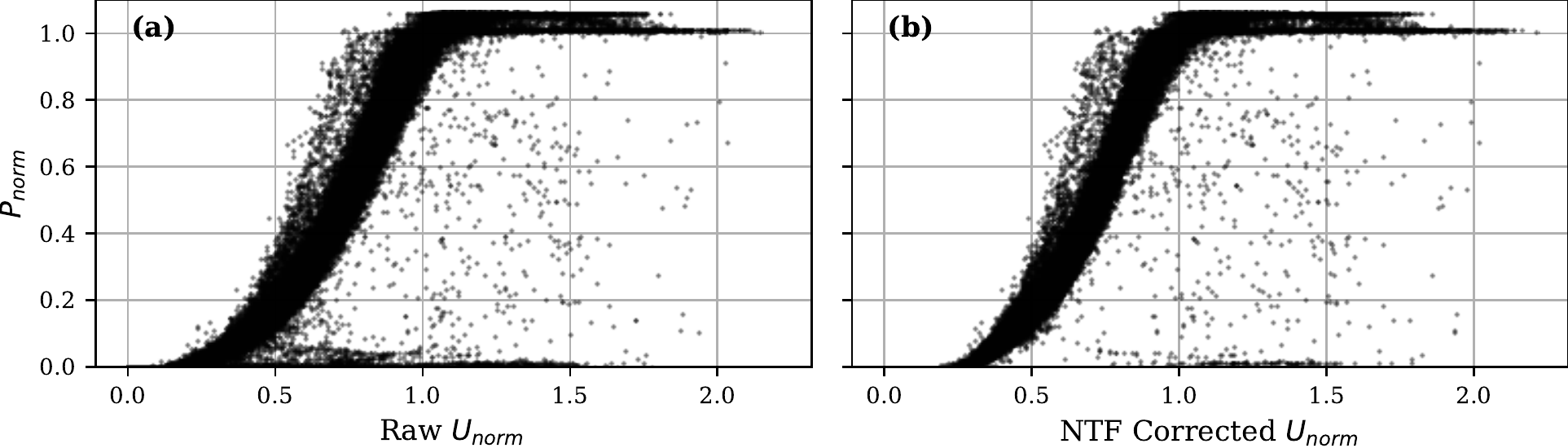}
    \caption{Power curve for turbine 06: (a) before and (b) after pre-processing has been applied.}
    \label{fig:wtpc_correction_effects}
\end{figure*}

Comparing the met-tower-measured wind speed with the reference wind speed calculated from the nacelle-transfer-function-corrected SCADA wind speed, the two data sets have a normalized mean absolute error (NMAE) of 7.79\%, which is calculated as:
\begin{equation}\label{eqn:NMAE_defn}
    NMAE=\frac{\sum_{i=1}^{N}|U_{true,i}-U_{pred,i}|}{\sum_{i=1}^{N}|U_{true,i}|},
\end{equation}
where \(N\) is the number of data points and \(U\) may be any parameter and is taken to be wind speed in this case. Figure \ref{fig:wtpc_correction_effects} illustrates the power curve for turbine 06 before (figure \ref{fig:wtpc_correction_effects}(a)) and after (figure \ref{fig:wtpc_correction_effects}(b)) the pre-processing is applied.

\section{Data Filtering}\label{sec:filtering}
After applying the pre-processing procedures discussed above, the SCADA data still contains many outliers due to several factors such as power curtailment, maintenance, off-design performance, or sensor fault. Figure \ref{fig:wtpc_correction_effects}(b) illustrates SCADA data available for turbine 06 after the nacelle transfer function and density corrections are applied. The need for appropriate filtering of these outliers is clear, especially for the training of ML models, as outliers in the training data may jeopardize the accuracy of the models. Regions of the WTPC that are not of interest to the analysis should also be removed so that the model is not diluted by these points. For these reasons, data filtering can be broken down into two steps: 1) region removal from the power curve and 2) outlier removal. These steps are described in the following subsections.

\subsection{Region Removal from the Power Curve}\label{sec:sub_region_removal}
As the eventual goal of this study is to investigate turbine performance and wake effects, only the most salient regions of the WTPC should be selected. Region one is the region where the average wind speed is below the cut-in wind speed of 3 m$\cdot$s$^{-1}$ and is characterized by primarily random fluctuations in power. Furthermore, since this region occurs very infrequently, it is removed. Additionally, region three, the region where the average wind speed is above the rated wind speed of 11 m$\cdot$s$^{-1}$, is not important for wake studies, since power capture is fixed at the rated power and the main turbine risks are associated with loads \cite{Howland2019}. This region is removed by rejecting points with an average wind speed above 13 m$\cdot$s$^{-1}$. This way, the transition from region two to region three is kept.

The turbines being studied have a rated power of 2.3 MW but are capable of boosting power output up to 2.5 MW if environmental conditions are acceptable and demand on the electricity grid is high. This power boost cannot be predicted, however, from environmental data alone, since it depends on grid conditions. For this reason, it is rejected by removing all points with power above 2325 kW. Finally, when the turbine rotor is being spun up to operating speed, the turbine may draw more power than it produces. If this is the case, the recorded power will be negative. These points are of no interest and dilute the models, so all points with power less than zero are rejected.

When training isolated-turbine models, the interest is the WTPC without considering wake effects. For this reason, wake conditions need to be removed from the turbine data sets according to the IEC standard \cite{IEC2017}. In contrast, when training ML models for use with the clustered-turbine approach, it is important to include wakes in the training data set, as the models will likely be interrogated under waked conditions. Thus, the filters considered should be capable of being applied to data sets both including and excluding waked sectors. If region and wake removal are applied, the average number of data points remaining across the 25 turbines is 18,677. Amongst the various turbines, the minimum number of remaining points is 4,331 and the maximum number is 34,727. This wide range in remaining points is to be expected since some turbines, such as the first row, are generally unwaked and will have many good points while other turbines, such as the middle row, are waked for almost all wind sectors and will not have many samples not affected by wakes. 

\subsection{Outlier Removal}\label{sec:sub_outlier_removal}
This study will consider four types of outlier filters to generate ``clean'' SCADA data for the training of ML models. The considered filters are binning, K-means clustering, GP, and DNN. These filters are applied to all 25 wind turbines and the data-rejection rates and final power curves will be analyzed to estimate the performance of the filters.

First, the binning filter is considered. This approach splits the power curve into 0.5 m$\cdot$s$^{-1}$-wide bins in wind speed \cite{IEC2017}. The mean power is calculated for each bin, as well as the standard deviation in power. Points with a power value further than 2 standard deviations from the mean power are rejected, with the choice of 2 standard deviations coming from a sensitivity analysis balancing apparent correct outlier removal with the number of rejected points.

Second, the K-means clustering filter uses the K-means algorithm to cluster data. For this work, 10 clusters are generated, as in the original paper \cite{Yesilbudak2018}, then the Mahalanobis distance from the respective cluster center is calculated for all points in each cluster. The Mahalanobis distance gives the distance between an observation vector of several dimensions, \(\bf{x}\), and the mean vector of all the observations being considered, \(\pmb{\mu}\), scaled using the covariance matrix \(\Sigma\):
\begin{equation}\label{eqn:m_dist_def}
    M\left(\bf{x}\right)=\sqrt{{\left(\bf{x}-\pmb{\mu}\right)}^T\Sigma^{-1}{\left(\bf{x}-\pmb{\mu}\right)}}.
\end{equation}
The covariance matrix, \(\Sigma\), is calculated uniquely for each cluster. All the SCADA parameters are considered except time. Once a Mahalanobis distance has been assigned to every data point, the mean and standard deviation of the Mahalanobis distance for each bin is calculated. Points further than 2 standard deviations from the mean value in each bin are rejected as in the original work.

The third filter considered is the automatic GP filter\cite{Manobel2018}, for which a single GP regression model is fit to the SCADA data using wind speed as the input and power as the output, with a different model being trained for every turbine. Since the GP model reports a standard deviation value for predicted points, upper and lower bounds on the power curve can be applied by adding 4 times the standard deviation curves to the GP-generated power curve, as prescribed in the original paper\cite{Manobel2018}. Points beyond these bounds are rejected and the model is re-trained on the remaining points. This procedure repeats until no more points are rejected.

The final filter considered is an automatic DNN filter, which operates on a similar principle to the GP filter, yet with a few differences. A DNN model of 5 fully connected dense layers of 150 neurons each is trained to predict power using the entire SCADA data set of each turbine using 20 epochs. These hyperparameters were chosen manually to combine good accuracy with low computational cost. Though hyperparameter optimization for this model was performed using the Deep Hyper package \cite{Romit2020}, the resultant optimized model significantly increased computational cost with only a marginal increase in accuracy. Thus, manually selected hyperparameters are used. A new model is trained for every turbine, then used to predict power for every point in the data set. The absolute differences between the true and predicted power of each point are ascribed as an error. The average error across the data set is calculated, as well as the standard deviation. All points with an error greater than 4 times the standard deviation plus the average error are rejected. This process is repeated using the previously filtered data as new input data. Once the filter rejects fewer than 75 points, the process terminates. The number of standard deviations and the rejection threshold was determined through a preliminary sensitivity analysis.

To be effective, the DNN filter needs to have an optimal selection of input parameters. To select the best combination of inputs, a simple brute force method is adopted. All of the SCADA parameters are considered candidates for inputs except for time, which is replaced by the time of the day. Each combination is required to have wind speed as an input. All combinations of the remaining parameters are considered and used to train the DNN model. Five-fold cross-validation is used, in which a data set is shuffled into five subsets and the model is tested on each one of the sets individually while being trained on the remaining four, to evaluate the performance of the model on each different input parameter combination \cite{Yang2007}. The NMAE is calculated averaging over the 5 folds, and the input parameter set with the lowest NMAE is selected. The standard deviation of NMAE scores is also reported, calculated over the 5 folds. Table \ref{tab:raw_dnn_input_list} lists the top 5 input combinations. This analysis was applied to turbine 06 with waked sectors removed from the data, as well as the regions prescribed in section \ref{sec:sub_region_removal}. The top results of the optimization are very similar in scores. Considering that the top 10 combinations all have NMAE scores below \(4\%\) and that all results are under \(6.5\%\), it is reasonable to conclude that model of this data set is relatively insensitive to input parameter selection, and thus the selection of inputs is not critical. For simplicity, and to marginally decrease computational cost, the inputs selected for the DNN filter are wind speed and $TI$.

\begin{table*}[b!]
    \centering
    \begin{tabular}{|c|c|c|c|}
        \hline
        \textbf{Combination ID} & \textbf{Parameters} & \textbf{NMAE [\%]} & \textbf{STD [\%]} \\ \hline
        1 & Wind Speed, $TI$, Temperature & 3.68 & 0.26 \\ \hline
        2 & Wind Speed, $TI$ & 3.72 & 0.25 \\ \hline
        3 & Wind Speed, $TI$, Time of Day, Temperature & 3.76 & 0.24 \\ \hline
        4 & Wind Speed, $TI$, Time of Day, Temperature, Density & 3.76 & 0.33\\ \hline
        5 & Wind Speed, $TI$, Time of Day, Temperature & 3.80 & 0.28 \\ \hline
    \end{tabular}
    \caption{Five best input-parameter combinations for the DNN filter trained on raw data.}
    \label{tab:raw_dnn_input_list}
\end{table*}

\subsection{Application of the Data Filters}\label{sec:sub_filter_application}
\begin{figure*}[b!]
    \centering
    \includegraphics[width=\textwidth]{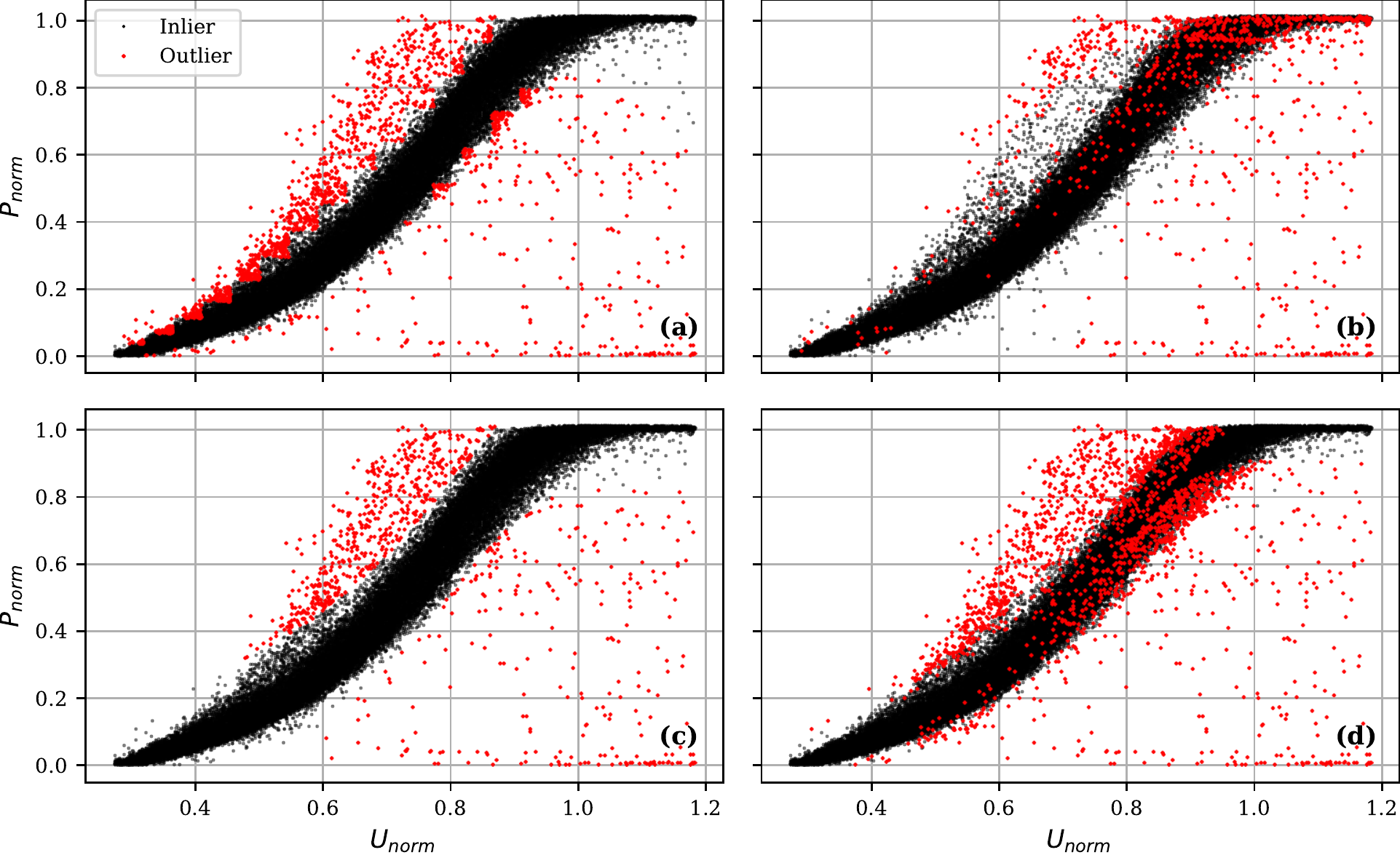}
    \caption{Application of data filtering for turbine 06 including waked sectors and using different methods: (a) binning, (b) K-means cluster, (c) GP, and (d) DNN.}
    \label{fig:filter_application}
\end{figure*}

The above-described four filters are applied to all 25 wind turbines to remove outliers. As mentioned above, each filter will be applied to the turbine data sets twice, once including wake conditions, and once excluding wakes, with region removal applied in all cases. Filters that do not consider $TI$ as an input, namely the binning and GP filters, are expected to perform worse when filtering data sets including wakes, as the increased variability in $TI$ due to wakes will create an increased variability in power not captured by using wind speed as the only input. Since the K-means clustering and DNN filters do accept $TI$ as an input, they are not expected to be adversely impacted by the inclusion of waked sectors. Figure \ref{fig:filter_application} shows the results of filtering when applied to turbine 06 and waked sectors are included. Table \ref{tab:filter_rejection_rates} reports the statistics on rejection rates for all the filters applied across all 25 wind turbines for both unwaked and waked input sets.
\begin{table}
    \centering
    \begin{tabular}{|c|c|c|c|c|c|c|}
        \cline{2-7}
        \multicolumn{1}{c|}{} & \textbf{Filter} & \textbf{Min} & \textbf{\(25^{th}\) \%-ile} & \textbf{Median} & \textbf{\(75^{th}\) \%-ile} & \textbf{Max} \\ \hline
        \multirow{4}*{\textit{Unwaked}} & Binning & 158 & 360 & 591 & 741 & 989 \\ \cline{2-7}
         & K-Means Clustering & 84 & 210 & 246 & 489 & 700 \\ \cline{2-7}
         & GP & 11 & 94 & 151 & 202 & 413 \\ \cline{2-7}
         & DNN & 15 & 219 & 353 & 648 & 1143 \\ \hline
        \multirow{4}*{\textit{Waked}} & Binning & 1321 & 1500 & 1550 & 1631 & 1789 \\ \cline{2-7}
         & K-Means Clustering & 734 & 787 & 836 & 883 & 1013 \\ \cline{2-7}
         & GP & 358 & 509 & 729 & 834 & 1071 \\ \cline{2-7}
         & DNN & 1150 & 1537 & 1642 & 1899 & 2389 \\ \hline
    \end{tabular}
    \caption{Statistics of rejected points by application of the filters to all 25 wind turbines.}
    \label{tab:filter_rejection_rates}
\end{table}
These results show that the binning filter is inefficient as it forms rectangles of acceptable points on the power curve, cutting off many good points, which also causes this filter to have high rejection rates. Both the K-Means clustering and the DNN filter appear to remove good points that fall on the main body of the power curve. Both of these filters, however, consider more parameters than simply wind speed and power, so they may be detecting outliers in other parameters. The K-Means filter struggles at higher wind speeds because the clusters in that region have fewer points, making filtering based on standard deviation less useful. Since the DNN filter does not use clustering, under the assumption that the DNN model is accurately capturing typical performance, it is reasonable to conclude that the apparent inliers that have been marked as outliers likely are outliers considering other parameters, not shown in figure \ref{fig:filter_application}. Finally, the GP filter performs well at trimming points far away from the main body of the WTPC, yet only considers wind speed and power, and so may fail to identify outliers in other parameters.

With these considerations in mind, the DNN filter is selected for use in this study. Though the rejection rates are often the highest out of the 4 considered filters, they are still reasonable considering the average data set size and do not threaten to reduce the number of points too greatly. Furthermore, it is one of the two filters that can detect outliers in parameters other than wind speed and power. Since the K-Means filter rejects points near the top of the power curve that no other filter rejects and leaves points near the bottom of the curve most other filters remove, the DNN filter functions better. Finally, since the goal is to improve ML models of turbine performance, it is reasonable to reject points that the DNN model struggles to predict so that these points do not dilute the training of future ML models. These points are removed from the data sets before training any model that predicts power. In later sections, the model predictions will be compared against both filtered and unfiltered SCADA data to illustrate the potential benefits of data filtering in assessing model performance.

\subsection{Data Filtering for Wind Speed and TI Models}\label{sec:sub_local_model_filtering}
\begin{figure*}[t!]
    \centering
    \includegraphics[width=\textwidth]{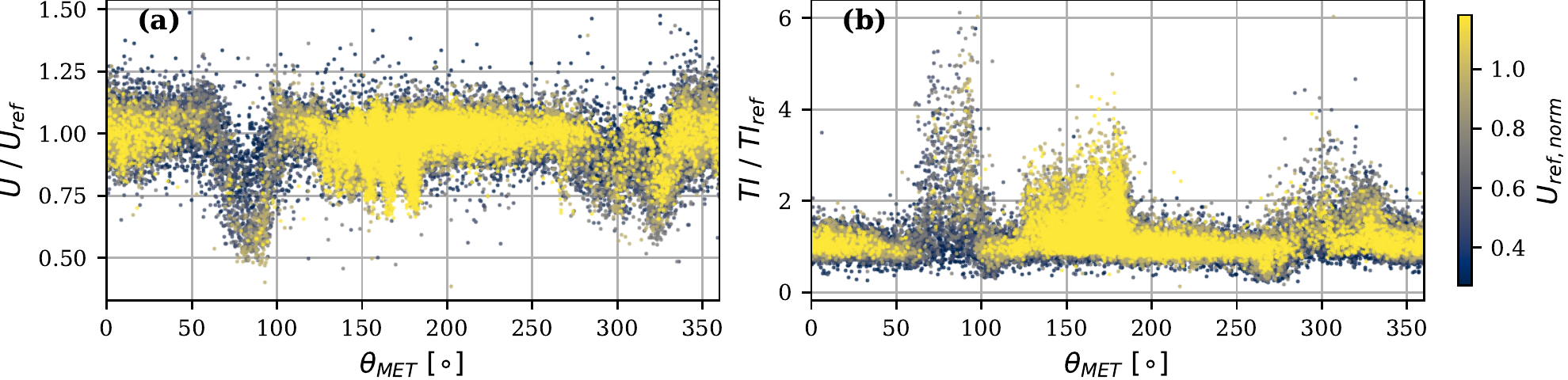}
    \caption{Local wind conditions for turbine 08 by varying wind direction: (a) Local wind speed; (b) $TI$.}
    \label{fig:reference_filter_plot}
\end{figure*}

Since the clustered-turbine modeling approach aims at predicting wind speed and $TI$ at each turbine location from reference conditions, it is reasonable to investigate if filtering should be applied to the training data for these models. Specifically, as will be discussed in section \ref{sec:modeling}, these models take reference wind speed, direction, and $TI$ as inputs and predict wind speed and $TI$ at the various turbine locations. The main source of error in these parameters is sensor fault, as these parameters are not susceptible to influences of the wind farm operator, grid demand, or other causes of outliers in power in SCADA data. Detecting these outliers, however, proves to be challenging due to the occurrence of wakes and speedups. For instance, figure \ref{fig:reference_filter_plot} illustrates the high variability due to wakes and speedups for turbine 08. The strong wake effects from turbines 09 and 07 are seen at wind directions of approximately 90$^\circ$ and 300$^\circ$, respectively. Additionally, the weaker wake effects from the first row of turbines can be seen within the wind sector between around 140$^\circ$ to 200$^\circ$, i.e. for southerly wind directions. These wake velocity deficits are accompanied by corresponding spikes in $TI$. While there are clear outliers, identifying these outliers is difficult, especially for $TI$, given its highly variable nature. Additionally, the topography, though it is relatively flat, may still introduce gradients in the flow for certain wind directions, resulting in slowdowns or speedups not explained by neighboring turbines. If there are any topographic gradients, points with this behavior should be included in the training data so that the models will account for these effects. Based on these considerations, data used for the wind speed and $TI$ models are left unfiltered to avoid removing important wake interactions or topographic effects, and in the expectation that, with enough good points, the models will be insensitive to scattered, random outliers. Thus, filtering will only be applied for the training data of turbine power models, whether isolated or clusters, and not for the wind speed and $TI$ models used in the clustered-turbine approach.

\section{Wind Turbine Modeling}\label{sec:modeling}
As mentioned above, two approaches to modeling are used: isolated- and clustered-turbine approaches. The purpose of the isolated-turbine approach is to aggregate data from either all turbines or a specific row of wind turbines, removing wakes, regions 1 and 3 of the WTPC, and outliers, to produce an ML surrogate for the WTPC that can vary with both wind speed and $TI$. On the other hand, the purpose of the clustered-turbine approach is to use a combination of wind speed, $TI$, and power models to capture environmental and turbine-to-turbine effects induced by neighboring turbines on individual turbines. These two approaches require the development of four different ML models: an aggregated turbine power model representative for the entire wind farm and individual turbine power models, individual turbine wind speed models, and individual turbine $TI$ models for each wind turbine. For each model category, an appropriate ML algorithm is selected and overall accuracy is discussed. When GP or SVM models are used, the Scikit learn package is used to generate these models \cite{Pedregosa2011}. When DNN models are used, the Keras package with the TensorFlow backend is used to generate the models \cite{Tensorflow2015}.

\subsection{Wind Turbine Power Models}\label{sec:sub_turbine_models}
Two types of turbine WTPC models are needed, a general WTPC model - used for the isolated-turbine approach - and an individual turbine WTPC model, which is to be generated for each turbine individually and is used in the clustered-turbine approach. A WTPC model needs to predict power output using selected SCADA parameters as inputs. As discussed in section \ref{sec:sub_outlier_removal}, the DNN model inputs are wind speed and $TI$. The DNN hyperparameters are identical to those used for the DNN filter. Other ML models will also be considered in addition to the DNN, such as GP and SVM models. The inputs to all three ML models will be kept identical. Since the goal is to predict turbine performance without the impact of the environment or wakes, the inputs to the model are wind speed and $TI$ at each turbine location rather than reference conditions. Waked wind sectors are also removed. Furthermore, the DNN filter discussed in section \ref{sec:sub_filter_application} is applied to remove off-design conditions. Filtering, wake removal, and region removal are applied to each turbine individually. Then, the selected turbine data are aggregated. Each ML model can then be trained over an aggregated data set. The power predictions will also be compared against the power predicted by a power curve fit using the IEC standard \cite{IEC2017}. Finally, a novel hybrid model is proposed that predicts turbine power using as inputs the predicted powers from a DNN, GP, and IEC model. An SVM is used for this model to map the predicted power from the three other models to predict more accurate turbine power. With this approach, systematic errors in the other models might be detected and corrected.

\begin{figure*}
    \centering
    \includegraphics[width=\textwidth]{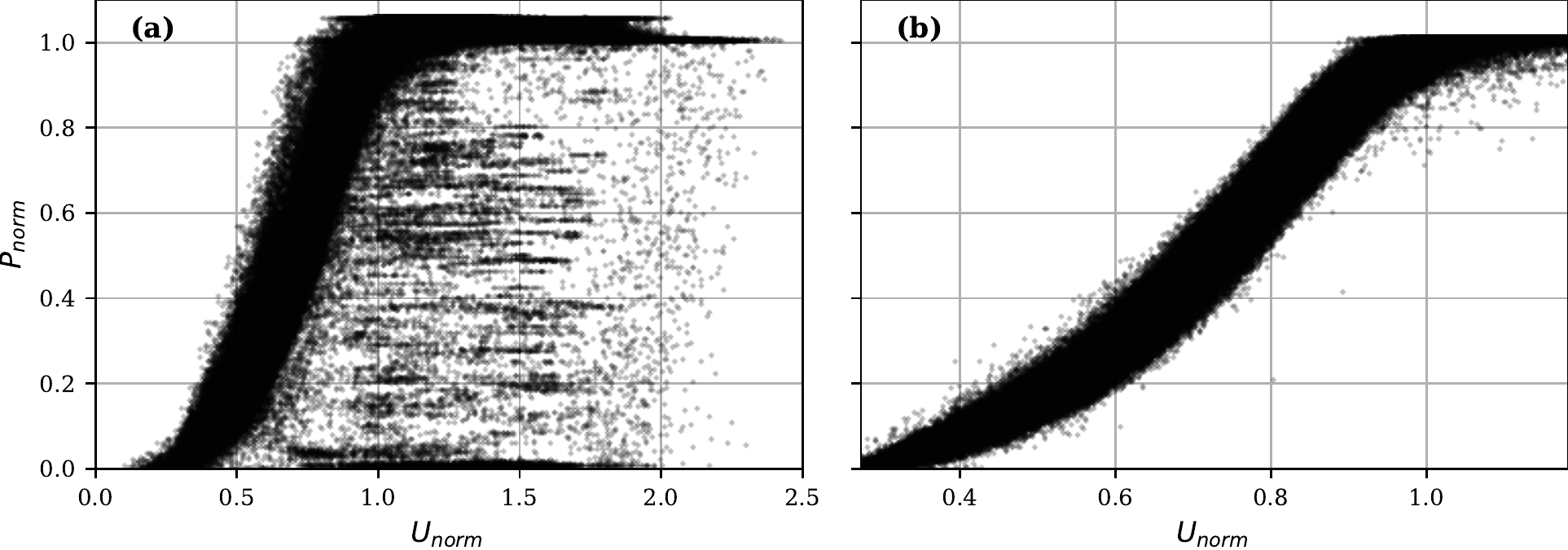}
    \caption{Training data set for the isolated-turbine model before (a) and after (b) region removal, wake removal, and DNN filtering.}
    \label{fig:isolated_turbine_training_data}
\end{figure*}

Two approaches for turbine data aggregation are used. The first considers all turbines in the farm simultaneously. Figure \ref{fig:isolated_turbine_training_data} shows the data set before and after the necessary pre-processing and filtering are applied. This corresponds to the most general turbine model for the farm. Additional row-specific models can be created by training the models using data aggregating along a specific row. Row 1 is the most southerly, row 2 is the middle one, and row 3 is the most northerly. Turbines 07, 73, and 74 are considered part of row 2 while turbines 75-78 are considered part of row 3. Following this approach, variations from row to row can be captured and investigated. Each ML model is to be evaluated on each approach to aggregation.

Five-fold cross-validation is used to evaluate each of the 5 ML models on each aggregation approach \cite{Yang2007}. Cross-validation allows the NMAE to be calculated for each fold, meaning a mean and standard deviation NMAE can be reported for each model. Table \ref{tab:isolated_turbine_model_evaluation} lists the results of the cross-validation. The hybrid model is the best in all row isolated cases as well as overall, so it is selected for use with the isolated turbine approach moving forward.
\begin{table*}[b!]
    \centering
    \begin{tabular}{|c|c|c|c|c|}
        \hline
         & \textbf{All} & \textbf{Row 1} & \textbf{Row 2} & \textbf{Row 3}\\ \hline
        \textit{IEC} & 5.35\% (0.01\%) & 5.40\% (0.03\%) & 5.11\% (0.02\%) &  5.44\% (0.03\%) \\ \hline
        \textit{GP} & 3.54\% (0.01\%) & 3.27\% (0.07\%) & 3.49\% (0.07\%) & 3.96\% (0.10\%) \\ \hline
        \textit{SVM} & 3.22\% (0.01\%) & 2.83\% (0.01\%) & 3.07\% (0.01\%) & 3.67\% (0.02\%) \\ \hline
        \textit{DNN} & 3.33\% (0.17\%) & 2.87\% (0.13\%) & 3.17\% (0.08\%) &  3.70\% (0.02\%) \\ \hline
        \textit{Hybrid} & \textbf{3.16\% (0.01\%)} & \textbf{2.76\% (0.01\%)} & \textbf{2.99\% (0.03\%)} & \textbf{3.60\% (0.03\%)} \\ \hline
    \end{tabular}
    \caption{Model scores evaluated on isolated-turbine data sets for all turbines and row-specific sets of turbines, mean NMAE with standard deviation of NMAE in parentheses.}
    \label{tab:isolated_turbine_model_evaluation}
\end{table*}

\begin{figure}
    \centering
    \includegraphics[width=0.5 \textwidth]{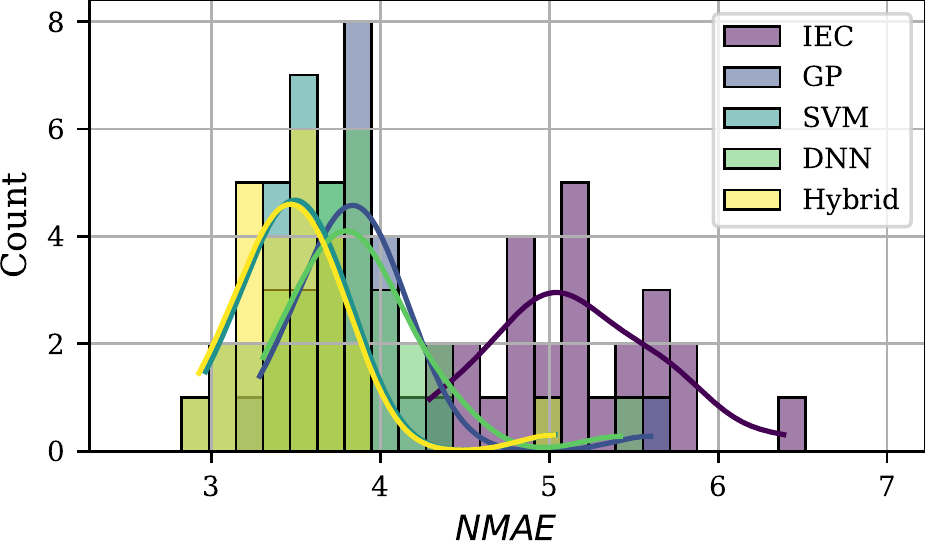}
    \caption{Error histogram for all five models applied to all 25 wind turbines using DNN filtered data including wakes.}
    \label{fig:turbine_model_error_hist}
\end{figure}

Next, the individual WTPC models needed for the clustered-turbine approach are considered. These models still consider as inputs just local turbine wind speed and $TI$ since the flow field modification introduced by clusters of turbines will be captured separately in wind speed and $TI$ models. In contrast to the isolated-turbine model, these models are trained over DNN-filtered data with wakes included since these models will be probed over waked sectors when simulating the farm as a whole. Five-fold cross-validation is again applied, now to each of the 25 turbines in the farm, for each of the five models under consideration identically to the isolated-turbine model case. The histogram of averaged NMAEs for each model over all 25 turbines is reported in figure \ref{fig:turbine_model_error_hist} while table \ref{tab:model_nmaes_filtered_waked} reports the statistics of averaged NMAEs for each model.

\begin{table}
    \centering
    \begin{tabular}{|c|c|c|c|c|c|}
        \hline
        \textbf{Model} & \textbf{Minimum} & \textbf{25\textsuperscript{th}\%-ile} & \textbf{Median} & \textbf{75\textsuperscript{th}\%-ile} & \textbf{Maximum} \\ \hline
        IEC & 4.29 & 4.88 & 5.12 & 5.50 & 6.39 \\ \hline
        GP & 3.28 & 3.67 & 3.85 & 4.02 & 5.60 \\ \hline
        SVM & 2.96 & 3.33 & 3.50 & 3.69 & 5.03 \\ \hline
        DNN & 3.30 & 3.63 & 3.85 & 4.03 & 5.42 \\ \hline
        Hybrid & \textbf{2.92} & \textbf{3.28} & \textbf{3.47} & \textbf{3.65} & \textbf{5.02} \\ \hline
    \end{tabular}
    \caption{Statistics on NMAE for various model types applied to all 25 wind turbines, trained on DNN filtered data [\%].}
    \label{tab:model_nmaes_filtered_waked}
\end{table}

From figure \ref{fig:turbine_model_error_hist}, the SVM and hybrid models appear almost indistinguishable. Likewise, the GP and DNN models are very similar. The IEC binning curve falls behind all other models in terms of accuracy. Considering table \ref{tab:model_nmaes_filtered_waked}, the SVM and hybrid models are the two leading models. In every case, the hybrid model produces slightly smaller errors than the SVM. In some applications, the increased training time of the hybrid model - since it must train a GP, SVM, and DNN model - may mean that the SVM is a suitable compromise between performance and computational cost. For this work, however, the hybrid model is selected for use with all the following applications of the clustered-turbine model.

\subsection{Wind Speed and TI Models for the Clustered-turbine Approach}\label{sec:sub_turbine_ws_ti_models}
Wind speed and $TI$ models are used in the clustered-turbine approach to capture all site-specific effects on the power performance of a turbine associated with its deployment within a specific cluster. For instance, these models should enable the detection of the intra-wind-farm variability of the wind field associated with the specific site topography and climatology, and the evolution and interaction of wind turbine wakes. This aim is accomplished by selecting the reference wind speed and $TI$ defined above, as well as wind direction from the met-tower, as inputs for the models, while the outputs are the wind speed and $TI$ at a given individual turbine location. In this way, all topographic and turbine-to-turbine effects are captured as well as the variation of these effects with reference wind speed, $TI$, and direction.

The inputs to the wind speed and $TI$ models could be determined by a brute force analysis similar to that used for the DNN model. Since these models will be used as the starting point to simulate clusters of turbines, however, it makes sense to select only the inputs that will be of interest to cluster simulations. These inputs will ultimately be supplied by the user to simulate a desired environmental condition. Thus, to keep the number of user-supplied inputs low, reference wind speed, $TI$, and direction are selected as the inputs to these models. Since each turbine wind speed is corrected with density normalization, the reference wind speed is also density-corrected. The nacelle transfer function corrects anemometer issues between turbines, so the averaging procedure can be done confidently.

There are two approaches to predicting turbine wind speed and $TI$. In the first approach, each parameter is predicted independently using a separate model trained to predict just that specific parameter. The models used are GP, SVM, and DNN. In the second approach, a DNN is trained to predict both wind speed and $TI$ simultaneously. Furthermore, when predicting $TI$, the inputs can be either reference conditions alone or reference conditions with the addition of predicted turbine wind speed. All of these approaches will be considered and the most accurate will be selected.

\begin{figure*}[b!]
    \centering
    \includegraphics[width=\textwidth]{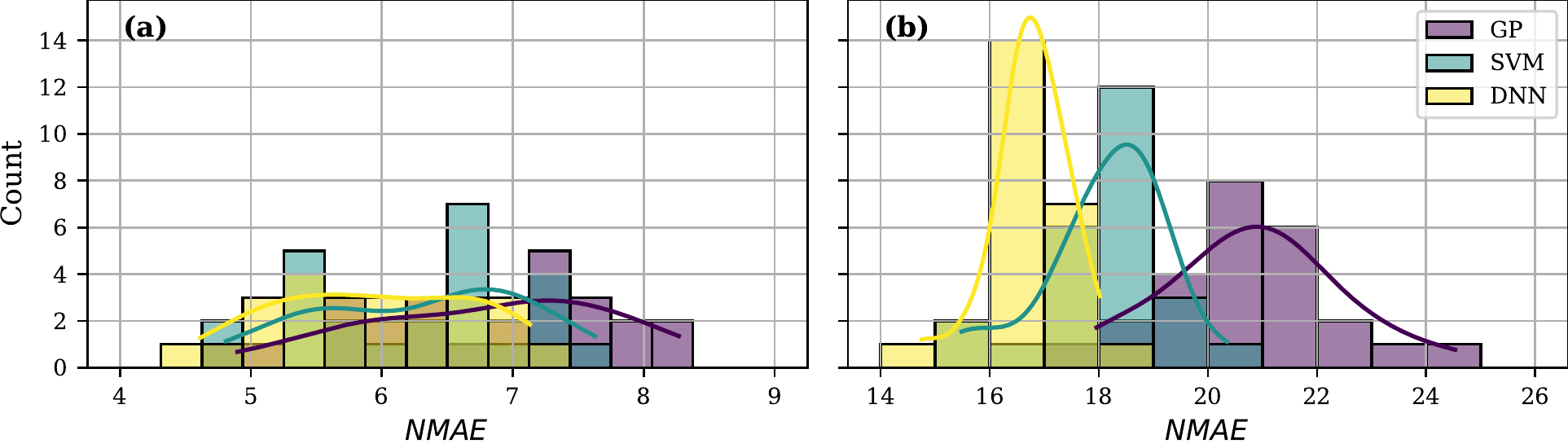}
    \caption{Error histogram for the models used with the clustered-turbine approach for all turbines predicting (a) wind speed and (b) $TI$ [\%].}
    \label{fig:local_model_ws_ti_error_hist}
\end{figure*}

The first approach to consider is predicting local wind speed from the reference conditions of wind speed, direction, and $TI$. These predictions are performed using a GP, SVM, or DNN model. Once again, the DNN hyperparameters are identical to the DNN filter and DNN turbine models. For each of the three models considered, an identical five-fold cross-validation approach is used as was used in the turbine model analysis in section \ref{sec:sub_turbine_models}. No filter is applied to the data before training as discussed in section \ref{sec:sub_local_model_filtering}. The error histogram for the three models is shown in figure \ref{fig:local_model_ws_ti_error_hist}(a) and the statistics are reported in table \ref{tab:model_nmaes_loc_ws}. From the results, it is seen that the DNN and SVM models perform very similarly, with the DNN performing very slightly better. Thus, we select the DNN as the model of choice for this application.

\begin{table}[b!]
    \centering
    \begin{tabular}{|c|c|c|c|c|c|}
        \hline
        \textbf{Model} & \textbf{Minimum} & \textbf{25\textsuperscript{th}\%-ile} & \textbf{Median} & \textbf{75\textsuperscript{th}\%-ile} & \textbf{Maximum} \\ \hline
        GP & 5.04 & 6.16 & 7.30 & 7.80 & 9.13 \\ \hline
        SVM & 4.96 & 5.68 & 6.76 & \textbf{6.94} & \textbf{7.76} \\ \hline
        DNN & \textbf{4.74} & \textbf{5.67} & \textbf{6.22} & 6.95 & 7.88 \\ \hline
    \end{tabular}
    \caption{Statistics on NMAE of predicted local wind speed for various model types applied to all 25 wind turbines.}
    \label{tab:model_nmaes_loc_ws}
\end{table}

The next approach to consider is predicting $TI$ directly from the reference conditions wind speed, direction, and $TI$. The approach is identical to the approach used to evaluate the wind speed models above. Similarly, the error histogram and statistics are reported in figure \ref{fig:local_model_ws_ti_error_hist}(b) and table \ref{tab:model_nmaes_loc_ti} respectively. In this case, the DNN model performs better than the other two models, though compared to the SVM model its performance improvement is often minimal. Though the NMAE values for these models are significantly higher than those for predicting local wind speed, given the increased difficulty in modeling and predicting turbulence, this is to be expected. Furthermore, power predictions are less sensitive to $TI$ than to local wind speed, so errors in $TI$ may have a smaller impact than errors in local wind speed.

\begin{table*}
    \centering
    \begin{tabular}{|c|c|c|c|c|c|}
        \hline
        \textbf{Model} & \textbf{Minimum} & \textbf{25\textsuperscript{th}\%-ile} & \textbf{Median} & \textbf{75\textsuperscript{th}\%-ile} & \textbf{Maximum} \\ \hline
        GP & 19.62 & 21.51 & 23.14 & 24.01 & 30.92 \\ \hline
        SVM & 15.91 & 17.97 & 18.87 & 19.19 & 20.94 \\ \hline
        DNN & 15.36 & 17.28 & 18.10 & 18.55 & 19.30 \\ \hline
        $TI$ from Local WS & \textbf{13.25} & \textbf{14.05} & \textbf{14.27} & \textbf{15.12} & \textbf{16.55} \\ \hline
    \end{tabular}
    \caption{Statistics on NMAE of predicted local $TI$ for various model types applied to all 25 wind turbines.}
    \label{tab:model_nmaes_loc_ti}
\end{table*}

$TI$ models could benefit from having the turbine wind speed as an input in addition to the reference conditions already considered. In practice, the wind speed model could first predict the turbine wind speed from the reference conditions, then the $TI$ model would predict turbine $TI$ from the reference conditions plus the predicted turbine wind speed. To assess this change of inputs, the previous approach is applied. Real turbine wind speed is provided as an input rather than predicted turbine wind speed for this analysis to avoid mixing sources of error. The results of this analysis are also listed in table \ref{tab:model_nmaes_loc_ti} and confirm that the latter approach outperforms all the other approaches.

Finally, it is considered whether predicting local wind speed and $TI$ simultaneously using a single model produces more accurate predictions than predicting each parameter individually. In this case, the inputs are reference wind speed, direction, and $TI$. A DNN is trained using these inputs to predict as output both the local wind speed and $TI$. As the information on the two conditions can influence the other in this modeling method, any interaction between the two conditions might be captured. The identical five-fold cross-validation analysis procedure is applied to this case as previously, modified to maintain the two-dimensional distribution of the outputs: turbine wind speed and $TI$. The resultant statistics in NMAE for both wind speed and $TI$ are reported in table \ref{tab:multi_model_nmaes}. While the accuracy is lower than other predictive methods, the model handled a prediction problem of increased complexity, so the scores are reasonable. The maximum scores are quite high, indicating that one turbine or a few turbines were not well captured by the model.

\begin{table*}
    \centering
    \begin{tabular}{|c|c|c|c|c|c|}
        \hline
        \textbf{Parameter} & \textbf{Minimum} & \textbf{25\textsuperscript{th}\%-ile} & \textbf{Median} & \textbf{75\textsuperscript{th}\%-ile} & \textbf{Maximum} \\ \hline
        WS & 5.14 & 6.31 & 7.12 & 8.35 & 100.72 \\ \hline
        $TI$ & 16.65 & 18.04 & 20.32 & 23.34 & 96.66 \\ \hline
    \end{tabular}
    \caption{Statistics on NMAE of predicted local wind speed and $TI$ using a single model trained to predict both parameters simultaneously.}
    \label{tab:multi_model_nmaes}
\end{table*}

Thus, after considering multiple approaches to predicting the wind speed and $TI$ and individual turbines as a function of the reference wind speed, direction, and $TI$, it is determined that using separate DNN models to predict each parameter produces the most accurate predictions. For best accuracy, turbine wind speed is to be predicted using a DNN model trained on reference wind speed, direction, and $TI$, while turbine $TI$ is to be predicted using a DNN model trained on reference wind speed, direction, and $TI$, as well as local turbine wind speed. When using these models to predict turbine wind conditions, first the wind speed model is used to predict the turbine wind speed. This output is then passed as an input, along with the reference conditions, to the $TI$ model to predict the turbine $TI$. With this approach, the wind speed and $TI$ at each turbine can accurately be predicted.

\subsection{A note on ML predictions as a function of Wind Direction}\label{sec:sub_wind_direction_wrapping}
\begin{figure*}[b!]
    \centering
    \includegraphics[width=\textwidth]{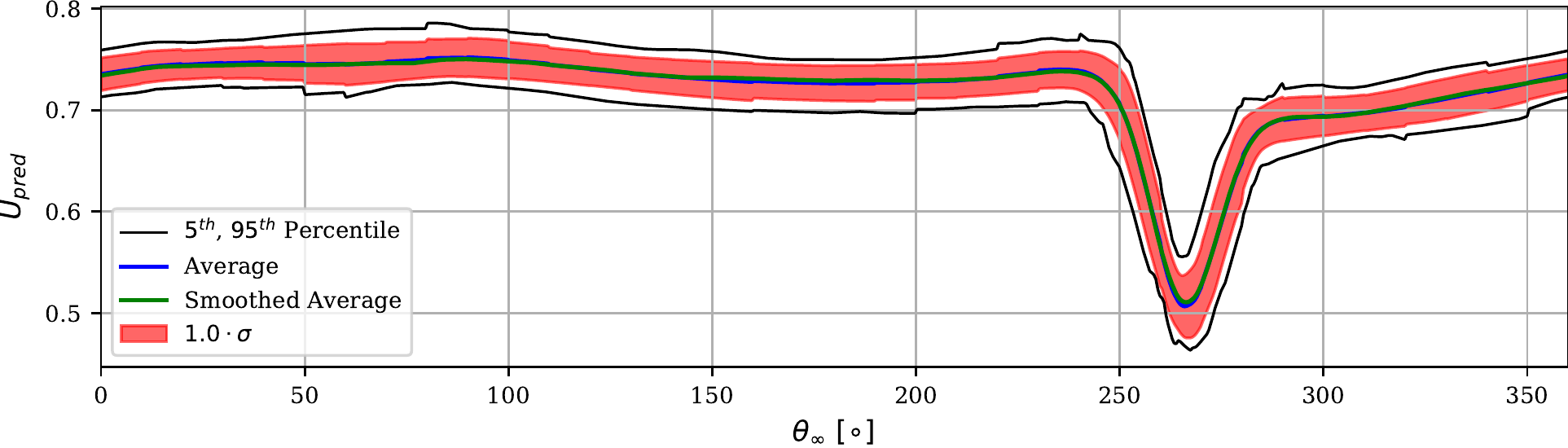}
    \caption{Filtering before smoothing multiple models to correct wind direction discontinuities.}
    \label{fig:filtered_avg_wd_shifted_models}
\end{figure*}

As a last step for the modeling procedure before considering the verification and validation of the models, a correction on the models with wind direction as input is proposed. While wind direction is an angular quantity, the ML models treat inputs as linear quantities. Thus, the model will extrapolate past 0$^\circ$ or 360$^\circ$ rather than interpolating. Additionally, although model behavior should be similar for angles around 0$^\circ$ and 360$^\circ$, the model treats the cases as extremes in wind direction. This creates a discontinuity at the $0^\circ$ wind direction. In other words, the boundary conditions in wind direction are expected to be periodic, but the ML models do not reproduce this behavior. To correct this discontinuity, an improved modeling approach is proposed. Several identical models are trained and then used to make individual predictions while adding to the wind direction an artificial angular shift. Practically, adding a shift before training moves the discontinuity by a set angular distance. The final prediction is then obtained as the average of the several predictions by removing the artificial angular shift. Thus, the individual discontinuities are averaged out. To further improve the predictions when predicting over continuous wind direction inputs, the average predictions can be smoothed using a Savitzky–Golay filter \cite{Schafer2011}. Before smoothing, however, the standard deviation is calculated, and all the points further than one standard deviation from the average prediction are rejected. The average curve is then recalculated from the remaining data and finally smoothed. These results are reported in figure \ref{fig:filtered_avg_wd_shifted_models} considering turbine 06. As a brief aside, the sharp dip in the predicted wind speed at around $270^\circ$ illustrates the model's capability to capture wake effects. From figure \ref{fig:layout_and_windrose}, turbine 06 is expected to be in the near wake of turbine 05 for westerly winds. Thus, the model is accurately capturing the expected phenomena. More details on predictions of wake interactions will be provided in section \ref{sec:model_validation}.

\section{Model Validation}\label{sec:model_validation}
Once the different modeling approaches have been defined and the models trained, model predictions can be investigated to ensure that the models perform as expected. Two analyses are performed for this purpose, i.e. for the model validation. First, the percent differences between turbine wind speed, $TI$, and power, all predicted through the clustered-turbine approach, and the reference wind speed and $TI$ as well as ideal power - the power produced by the given turbine operating in freestream conditions - are investigated for individual turbines over varying reference wind speeds, $TI$ values, and directions. This analysis highlights the capability of the clustered-turbine models to capture wind-farm effects, such as wake interactions decreasing local wind speed, increasing local $TI$, and decreasing power capture. More complex behaviors are also captured, such as local increases in wind speed and damping of $TI$ connected with speedup conditions, which occur for wind sectors adjacent to those associated with wake interactions \cite{Letizia2022(b)}. For the second analysis, power curves are generated using isolated-turbine models trained on data aggregated over either the entire wind farm or over specific rows of turbines. This analysis highlights the variability of the power curve to changes in $TI$ and turbine location, specifically among different turbine rows.

\subsection{ Variation of the Local Wind Condition and Power Capture}\label{sec:sub_local_variations}
To highlight wake impacts such as slowdowns or speedups, $TI$ increases, or power reductions, the variations of predicted parameters with respect to reference parameters are considered. First, the turbine wind speed model is investigated. As turbines 07 and 08 have the most interesting and complex wake interactions of the wind farm under investigation, they are chosen for this analysis \cite{El-Asha2017}. For each turbine, a synthetic input set is generated where the reference wind direction varies continuously within the range $0^\circ-360^\circ$, while the reference wind speed and $TI$ are held constant. The constant values used for wind speed vary from 3 m$\cdot$s$^{-1}$ up to 13 m$\cdot$s$^{-1}$ with 5 evenly spaced steps. The $TI$ values used are the 10\textsuperscript{th}, 25\textsuperscript{th}, 50\textsuperscript{th}, 75\textsuperscript{th}, and 90\textsuperscript{th} percentile values of reference $TI$ after removing wind speeds beyond the considered wind speed range. The percentage difference between the predicted local wind speed and the reference wind speed is reported for each synthetic data set to identify speedups or slowdowns. Additionally, the turbine SCADA data is binned in reference wind speed, direction, and $TI$ such that the bins are non-overlapping and centered on the wind speed and $TI$ values used in the analysis, while the wind direction bins are $10^{\circ}$ wide. If a given bin has fewer than 20 points, the results in the figures are masked for that sector under the supposition that the ML models do not have enough data to accurately reproduce the behavior in that sector.

\begin{figure*}[b!]
    \centering
    \includegraphics[width=\textwidth]{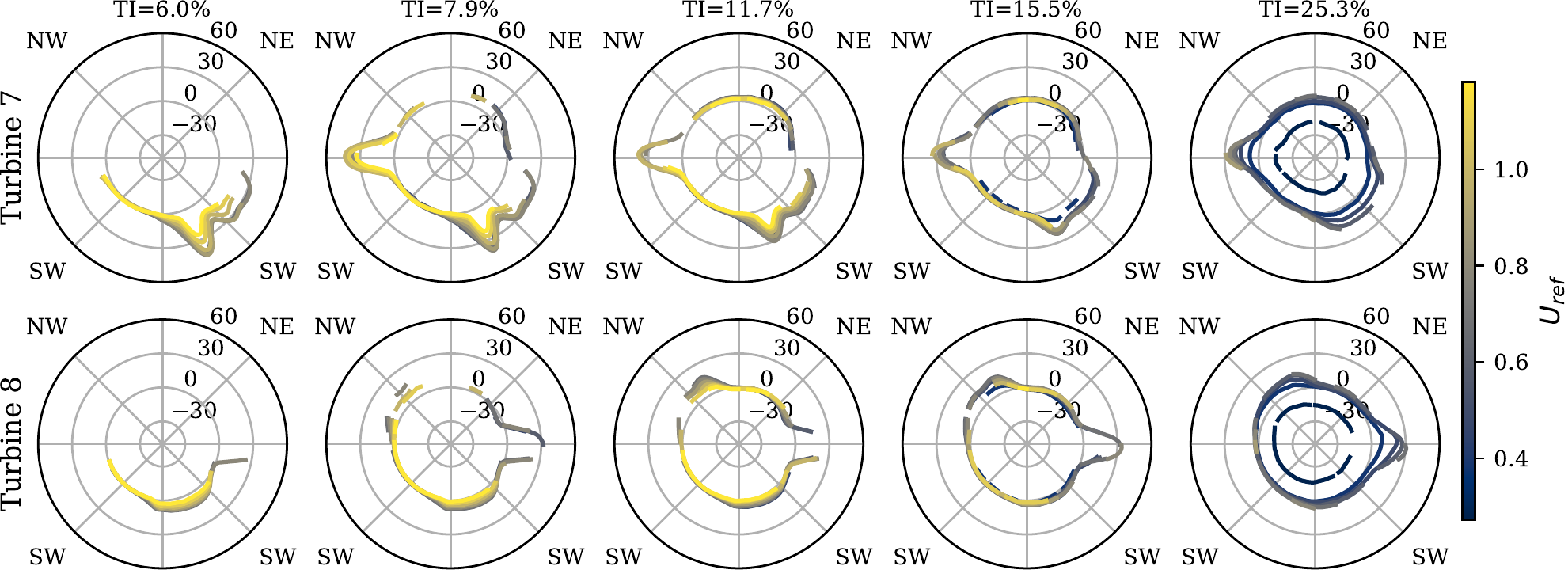}
    \caption{Local wind speed percentage losses for turbines 07 and 08. The analysis is performed for 10\(^{\circ}\)-wide bins in wind direction and rejecting sectors with fewer than 20 points. $TI$ increases from left to right with turbine 07 results in the top row and turbine 08 results in the bottom row.}
    \label{fig:ws_losses_turb_7_8}
\end{figure*}

In figure \ref{fig:ws_losses_turb_7_8}, slowdowns due to wakes for turbines 07 and 08 are evident from the regions with a positive percentage loss. Considering turbine 07, sharp increases in percentage variation are seen around 160\(^{\circ}\) and 270\(^{\circ}\), with a broader increase from roughly 100\(^{\circ}\) to 135\(^{\circ}\). Referring to figure \ref{fig:layout_and_windrose}(b), the two major peaks are expected due to interactions with turbines 73 and 08. Furthermore, the broad increase is also expected and is due to the second row of turbines, i.e. turbines 09 through 14. The magnitude of the slowdowns due to wakes is also consistent with other field studies performed for this wind farm \cite{El-Asha2017}. As $TI$ increases, the wakes become less prominent, and the peaks recede \cite{Zhan2020, Iungo2014}. Eventually, the wake effects from the first and second rows become very diffuse and extend over much of the region between 90\(^{\circ}\) and 180\(^{\circ}\). Finally, for some wind directions, the percent difference drops below zero, i.e. a speedup, which is expected to occur between wakes as mass conservation requires a region of faster-moving fluid between two regions of slower-moving fluid \cite{Letizia2022(b)}. For all wind directions at higher $TI$, however, the result for the lowest wind speed of 3 m$\cdot$s$^{-1}$ stays around -30\% to -60\%. Since this does not vary with wind direction and therefore does not vary with wake interaction, and since the magnitude is greater than expected for speedups \cite{Letizia2022(b)}, it can be concluded these results are not physical. This unexpected behavior at low wind speeds is likely not the result of boundary issues in the ML model, as the wind speed models are trained over all wind speeds, including wind speeds from 0 to 3 m$\cdot$s$^{-1}$, rather a real characteristic of the data set under investigation, which is going to be further investigated for turbine 08.

Repeating this analysis for turbine 08, the results are consistent with those of turbine 07. The percentage differences show a strong increase at 90\(^{\circ}\) due to turbine 09. More diffuse increases can be observed at about 315\(^{\circ}\), though at a $TI$ of 15.5\% two clear bumps emerge due to turbines 07 and 73. Additionally, increasing $TI$ causes the same smoothing out of wake losses as observed for turbine 07. Small speedup regions occur near the 90\(^{\circ}\) wake. For high $TI$ cases the same non-physical negative difference occurs at the lowest wind speed. To further investigate this behavior, the variation of model error with wind direction is calculated for turbines 07 and 08 using the absolute error and percentage error. The absolute error is given simply as \(|U_{pred}-U_{real}|\) and is always positive. The percentage error is given below in equation \ref{eqn:percent_error} and is signed.

\begin{equation}
    Percentage\ Error = \frac{U_{real}-U_{pred}}{U_{real}} \cdot 100\%
    \label{eqn:percent_error}
\end{equation}

\begin{figure*}[b!]
    \centering
    \includegraphics[width=\textwidth]{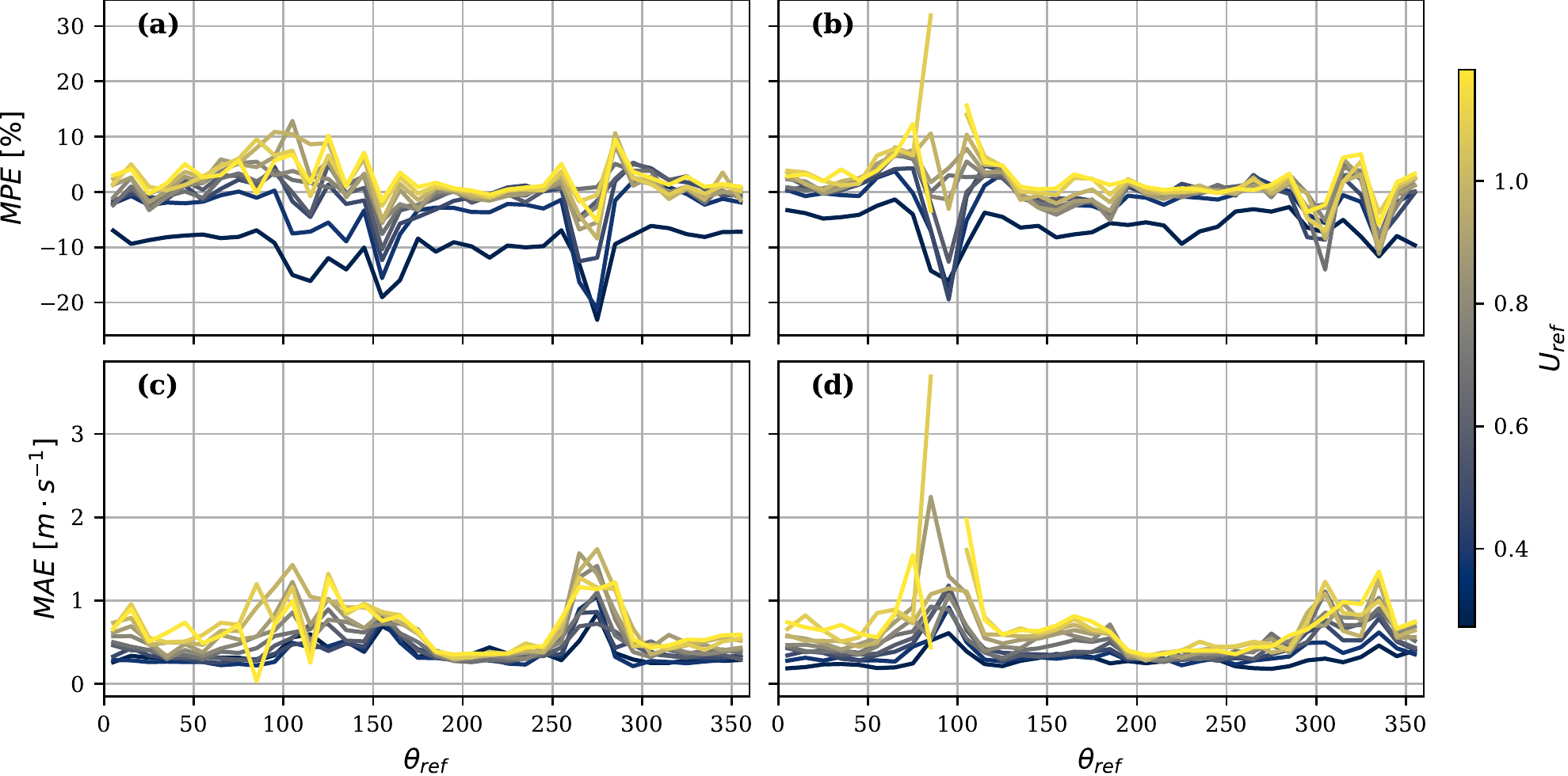}
    \caption{Model error varying with wind direction for turbine 07 mean percentage error (MPE) (a) and mean absolute error (MAE) (c) and turbine 08 MPE (b) and MAE (d).}
    \label{fig:err_vs_wd_turb_7_8}
\end{figure*}

The errors are binned in wind speed using 1 m$\cdot$s$^{-1}$-wide bins from 3 m$\cdot$s$^{-1}$ up to 13 m$\cdot$s$^{-1}$ and in wind direction using 10\(^{\circ}\)-wide bins from 0\(^{\circ}\) up to 360\(^{\circ}\). If a bin has fewer than 20 points, it is not considered in the analysis. The average absolute and percentage errors for each bin are reported considering turbines 07 and 08 in figure \ref{fig:err_vs_wd_turb_7_8} and show the strong effect of wakes on accuracy. Both turbines have spikes in waked regions in both absolute and percentage error, which can be interpreted as follows. First, wakes introduce greater turbulence variations, making predictions more difficult. Further, and more importantly, given the layout of the farm, strong wake interactions from directly neighboring turbines typically fall in the East-West direction. Yet, as the windrose in \ref{fig:layout_and_windrose}(a) indicates, the site is dominated by North-South wind. Thus, the strong wake regions also have low data availability, which means the DNN models will struggle to accurately capture behavior in these regions.

\begin{figure*}[b!]
    \centering
    \includegraphics[width=\textwidth]{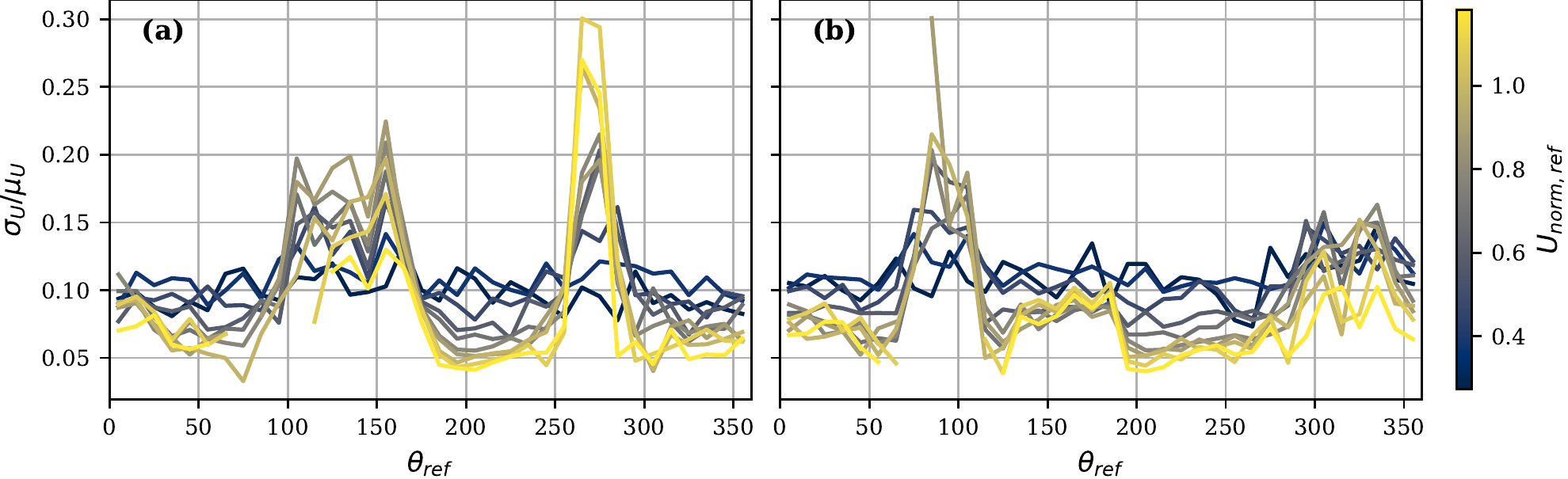}
    \caption{Variability of the SCADA data varying with wind direction for turbine 07 (a) and turbine 08 (b).}
    \label{fig:low_ws_data_variability}
\end{figure*}

Even outside of wake regions, the low wind speed results consistently have a larger magnitude of percentage error than other wind speeds, and the sign is always negative. An exception could be around 100\(^{\circ}\) for turbine 8, in figure \ref{fig:err_vs_wd_turb_7_8}(b) and (d), where the high wind speeds also seem to spike in the magnitude of percentage error, but in the opposite direction as the low wind speeds. As is shown in the graph, though, some of these bins fail to meet the data availability criterion, so this result could be attributed to low data availability. Since the low-wind-speed speedup trend is consistent for both turbines and all wind directions and is seemingly unaffected by data availability, it is reasonable to conclude that the model is reproducing a real feature in the data. Indeed, if this sector of the output has high variability in the real data, then the model could struggle to accurately predict turbine wind speed. To check this, following the same binning procedure used for figure \ref{fig:err_vs_wd_turb_7_8}, turbine 07 and 08 data is binned and the standard deviation of turbine wind speed normalized by the mean of turbine wind speed is plotted against the reference wind direction in figure \ref{fig:low_ws_data_variability}. For low wind speeds, the variability of turbine wind speed is high relative to the mean wind speed, except for waked regions, which can be disregarded for this analysis. This higher variability would make it more difficult for the model to extract features from the data and would contribute to increased errors, as has been shown. While causes for this variability could be further speculated, to avoid any issues this highly variable region may cause, the wind speed range is restricted between 5 and 13 m$\cdot$s$^{-1}$ for the remainder of the analyses in this section.

\begin{figure*}
    \centering
    \includegraphics[width=\textwidth]{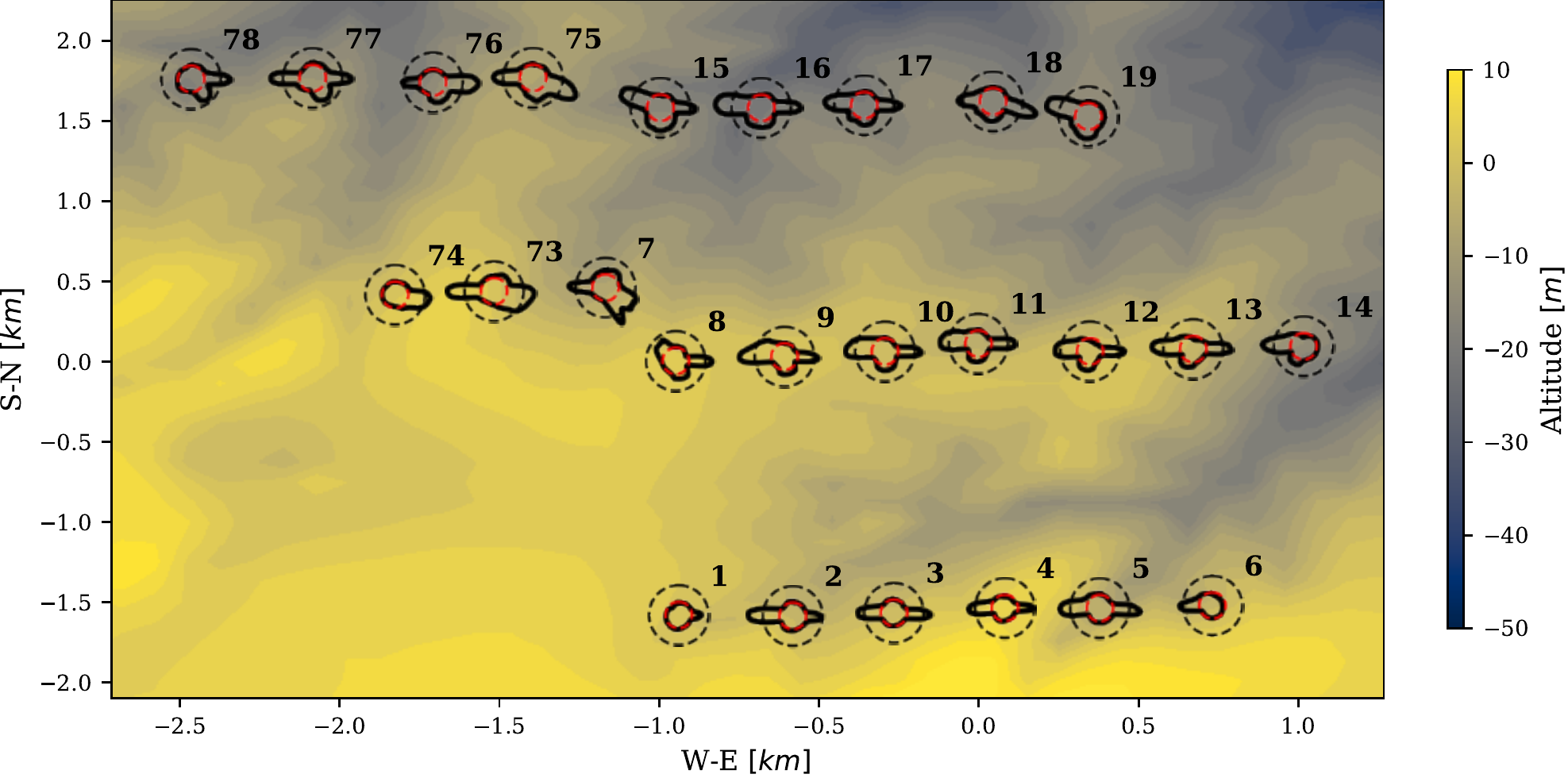}
    \caption{Wind speed percentage losses for the entire farm for a freestream condition with U=8 m$\cdot$s$^{-1}$ and \(TI=12\%\). Red dashed lines indicate 0\% change and the dashed black line indicates 25\% loss.}
    \label{fig:farm_ws_losses}
\end{figure*}

To demonstrate wind speed losses at the farm level, the reference wind speed is set to be 8 m$\cdot$s$^{-1}$ and the reference $TI$ is set to the median $TI$, roughly 12\(\%\). The foregoing procedure is applied to each turbine to generate wind speed losses for the specific environmental setting. The results are illustrated in figure \ref{fig:farm_ws_losses}. As can be noted, the wakes are all visible and pointing in the expected directions.

\begin{figure*}
    \centering
    \includegraphics[width=\textwidth]{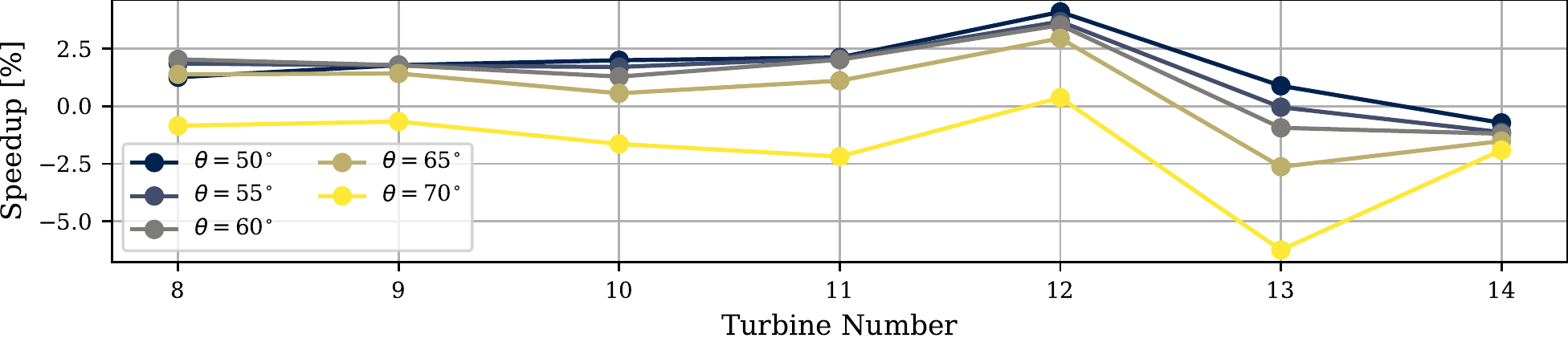}
    \caption{Speedup effects for the middle row of turbines at a reference wind speed of 8 m$\cdot$s$^{-1}$ and $TI$ of 8\%, varying wind direction.}
    \label{fig:row_speedups}
\end{figure*}

Figure \ref{fig:row_speedups} illustrates the speedup effect on the middle row of turbines (turbines 08 through 14) for a reference wind speed of 8 m$\cdot$s$^{-1}$ and $TI$ of 8\%. The varying wind direction highlights how wakes from the northerly row and even the middle row might combine to produce speedup effects. Interestingly, the turbine with the strongest speedup effect is turbine 12 and does not risk being waked from the northerly row for the wind directions considered. Once turbine 08 is no longer waked from above, it also begins to see strong speedup effects.

\begin{figure*}
    \centering
    \includegraphics[width=\textwidth]{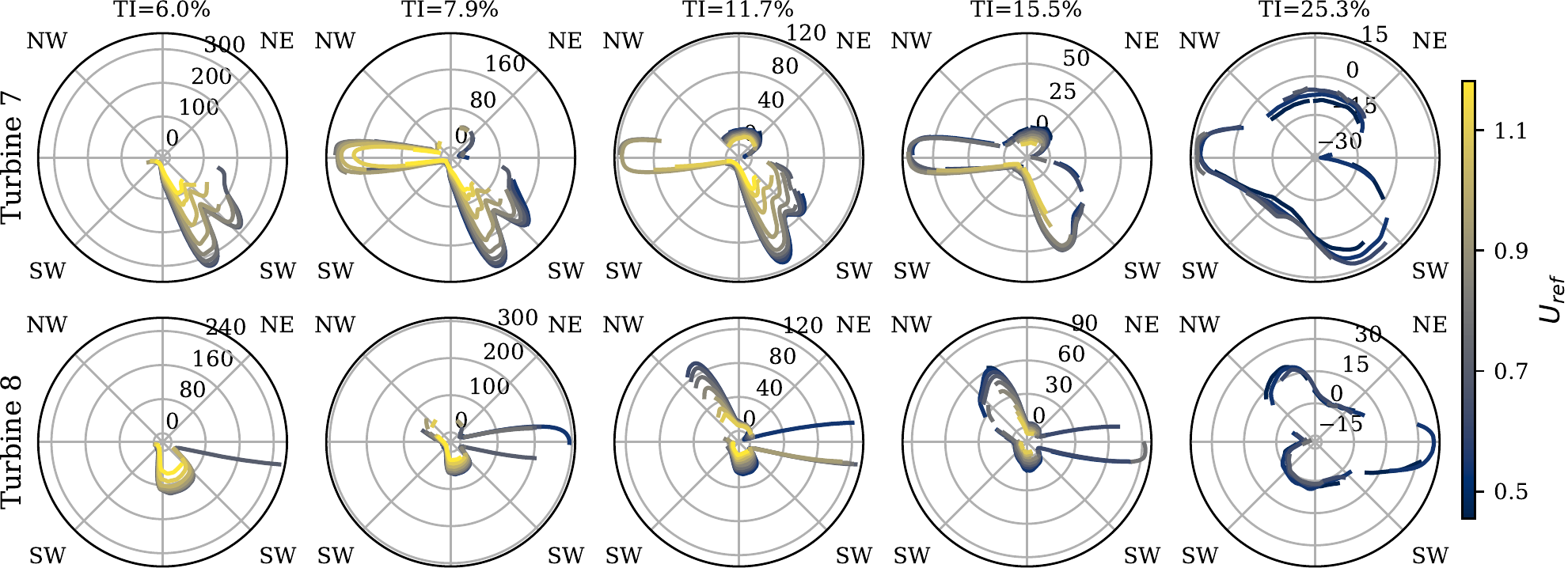}
    \caption{Local $TI$ percentage gains for turbines 07 and 08. The analysis is performed for 10\(^{\circ}\)-wide bins in wind direction and rejecting sectors with fewer than 20 points. Reference $TI$ increases from left to right with turbine 07 results in the top row and turbine 08 results in the bottom row.}
    \label{fig:combined_ti_gains}
\end{figure*}

\begin{figure*}
    \centering
    \includegraphics[width=\textwidth]{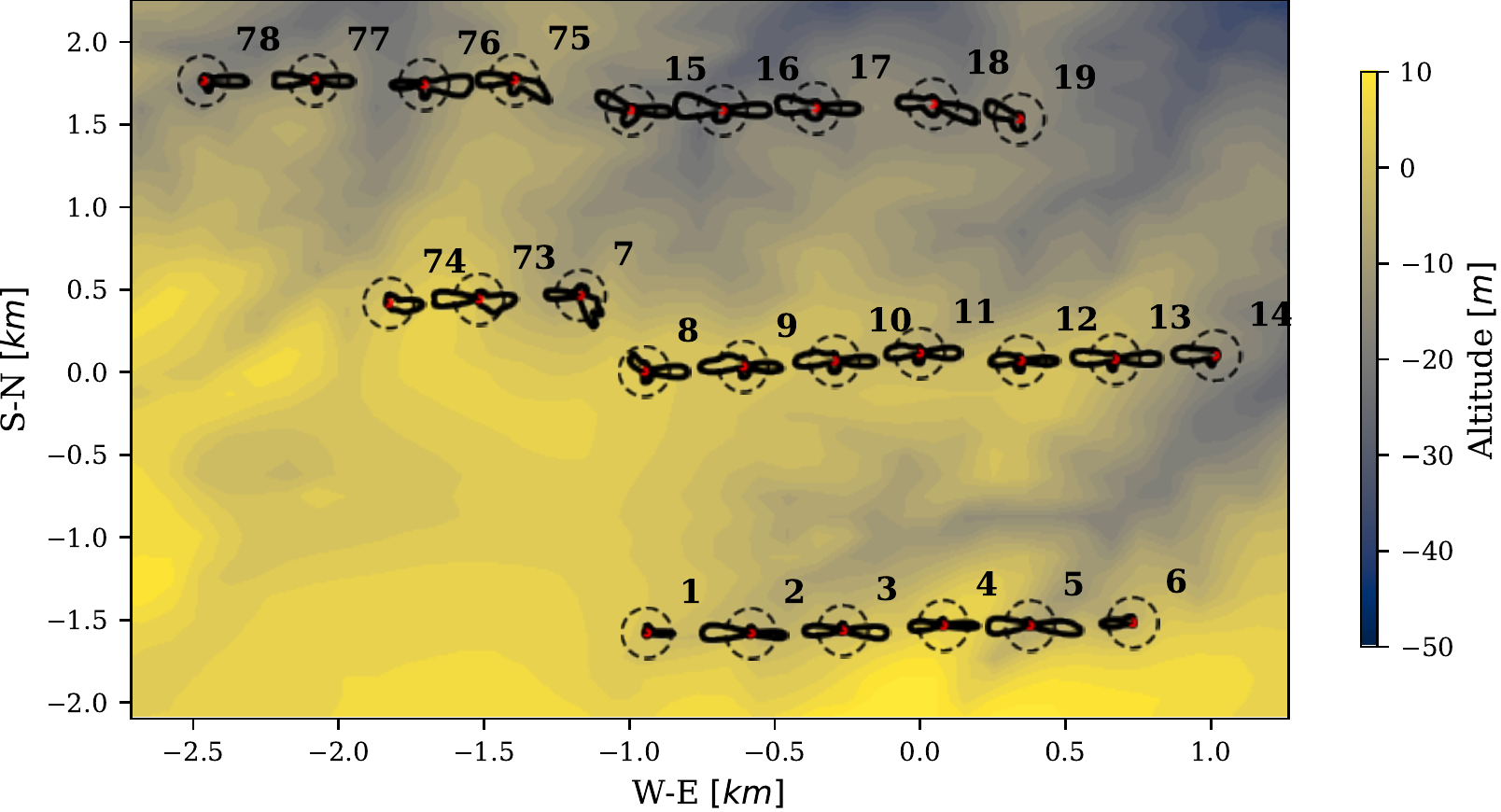}
    \caption{$TI$ percentage gains for the entire farm for a freestream condition with U=8 m$\cdot$s$^{-1}$ and \(TI=8\%\). Red and black dashed lines indicate 0\% and 150\% increase, respectively.}
    \label{fig:farm_ti_gains}
\end{figure*}

The local $TI$ models are validated following a similar procedure. In this case, the expected behavior is that $TI$ will spike in waked regions while remaining unaffected in unwaked regions. Figure \ref{fig:combined_ti_gains} illustrate this analysis for turbines 07 and 08, respectively. Note that since turbine wind speed is predicted and then used as an input to the local $TI$ models, the previous restriction of the wind speeds considered applies here. Thus, the minimum wind speed is restricted to 5 m$\cdot$s$^{-1}$. To observe the wake effects on $TI$ for the entire farm, the above analysis is applied to all turbines with the reference wind speed being set to 8 m$\cdot$s$^{-1}$ and the reference $TI$ being set to \(8\%\). The results are displayed in figure \ref{fig:farm_ti_gains}.

Once again, strong wake effects are seen, now in terms of $TI$ increases. The wake peaks are even clearer in this analysis than in the wind speed analysis. Furthermore, the previously observed trend of wake effects decreasing in strength as reference $TI$ increases is again noted here. The notable behavior of the turbine $TI$ models is in $TI$ damping. Therefore, this analysis suggests that the wake-generated turbulence intensity decreases with reducing reference $TI$ as a result of the reduced wake velocity deficit and, thus, transverse shear, which is the source for mechanical production of turbulence \cite{Pope2000}.

For both turbines 07 and 08, small regions exist next to waked regions where the $TI$ is decreased relative to the reference $TI$. For turbine 07, this is seen in the sector between 45\(^{\circ}\) and 90\(^{\circ}\), while for turbine 08, this is seen weakly on both sides of the wake at 90\(^{\circ}\) for moderate reference $TI$ values and again in the 270\(^{\circ}\) direction for high $TI$. This damping is expected because the speedup regions increase the mean wind speed without increasing the variability of wind speed, and, thus, $TI$ decreases.

\begin{figure*}
    \centering
    \includegraphics[width=\textwidth]{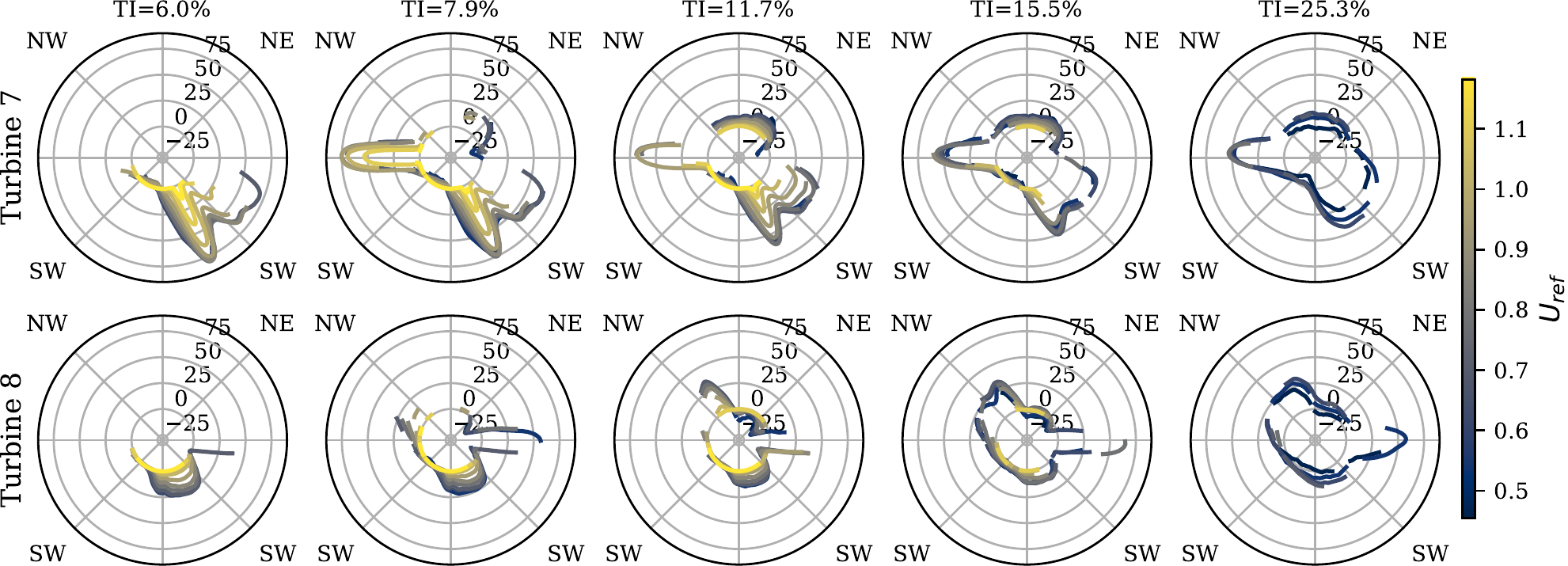}
    \caption{Local power percentage losses for turbines 07 and 08. The analysis is performed for 10\(^{\circ}\)-wide bins in wind direction and rejecting sectors with fewer than 20 points. Reference $TI$ increases from left to right with turbine 07 results in the top row and turbine 08 results in the bottom row.}
    \label{fig:combined_power_losses}
\end{figure*}

Finally, the behavior of turbine power can be investigated through the clustered-turbine approach. In previous analyses, local percentage variations were calculated by comparing local environmental parameters against supplied reference parameters. To obtain a reference power to compare against, the isolated-turbine model is used. When considering a given turbine, the isolated-turbine model selected should be trained on data aggregated across the row of the turbine under consideration. Since power production could potentially vary between the rows, it is not advisable to use the general isolated-turbine model trained on data aggregated across the entire farm. Doing this could lead to row power variation being confused with wake power variation. The results of this analysis for turbines 07 and 08 are reported in figure \ref{fig:combined_power_losses} and show power losses connected with wake interactions. Furthermore, the magnitude of the power losses is close to the values estimated through a more classical statistical approach for the same wind farm \cite{El-Asha2017}. Wind speeds above the turbine rated wind speed have almost no wake losses and lie in region 3 of the power curve. Additionally, wake losses decrease in magnitude with increasing reference $TI$, as thoroughly documented in the literature \cite{Zhan2020, Iungo2014, Iungo2016}. The model can capture small details in the variability of power performance, such as power increases due to a combination of local wind speed and $TI$ variation. For instance, turbine 07 typically exhibits a power boost between 45\(^{\circ}\) and 90\(^{\circ}\), while turbine 08 shows stronger boosts on both sides of the wake at 90\(^{\circ}\). Interestingly, these boost regions seem to align with $TI$ damping regions. While power boost regions are difficult to utilize to increase power production as they are small compared to wake regions, always occur next to wakes, and have much smaller magnitudes than wakes so that any positive effects are outweighed when considering long-term performance, the ability to predict boost regions is a step forward in understanding complex turbine wake interactions and improving wind farm control. Finally, considering the farm as a whole, similar results are obtained to the wind speed and $TI$ cases above, which are not presented here for brevity.

\subsection{Power Curve Modeling}\label{sec:sub_power_curve_modeling}
To analyze the isolated-turbine model, power curves are generated for various environmental conditions. The basic procedure to generate a power curve using an ML model is to generate a synthetic data set which is then used as input to a power model. This data set has a fixed value in all input parameters except wind speed and varies linearly in wind speed between a lower and upper limit. The resulting predicted power is the turbine power curve. When different values are used for $TI$, the variability of the power curve to $TI$ can be investigated.

\begin{figure*}[b!]
    \centering
    \includegraphics[width=\textwidth]{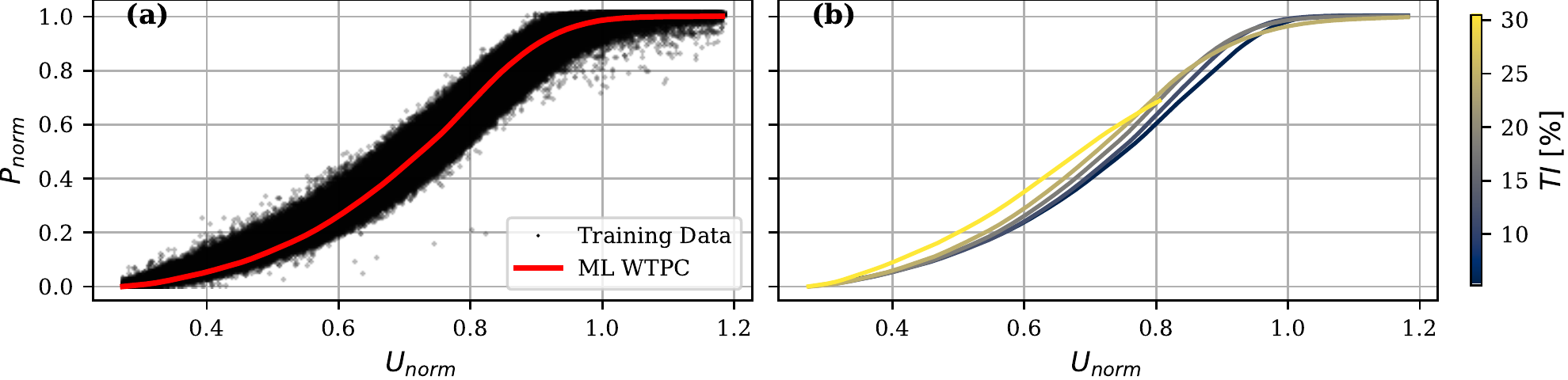}
    \caption{Power curves generated using the isolated-turbine model aggregating all turbines: (a) Training data set and power curve obtained with the median $TI$ from the training set (12\%); (b) ML-generated power curves obtained by varying the reference $TI$.}
    \label{fig:sample_pwr_curve}
\end{figure*}

First, the variability of power curves to $TI$ is investigated using the isolated-turbine modeling approach trained on data aggregated over the entire farm. Subsequently, using isolated-turbine models trained on data aggregated over individual rows, the sensitivity of different rows to $TI$ is investigated. The generated synthetic data set uses a constant value of reference $TI$ and a linearly varying wind speed from 3 m$\cdot$s$^{-1}$ up to 13 m$\cdot$s$^{-1}$. The model aggregating across the entire farm is used to generate the power curves in figure \ref{fig:sample_pwr_curve} where figure \ref{fig:sample_pwr_curve}(a) uses constant median reference $TI$ while figure \ref{fig:sample_pwr_curve}(b) generates varying curves using 5 evenly spaced percentiles in reference $TI$ from the 5\textsuperscript{th} percentile to the 95\textsuperscript{th} percentile, in this case, 5.2\% to 30.5\%. These curves are quality controlled using 0.5 m$\cdot$s$^{-1}$-wide bins in wind speed and 1\% wide bins in $TI$ and requiring bins to have at least 10 points per bin.

\begin{figure*}
    \centering
    \includegraphics[width=\textwidth]{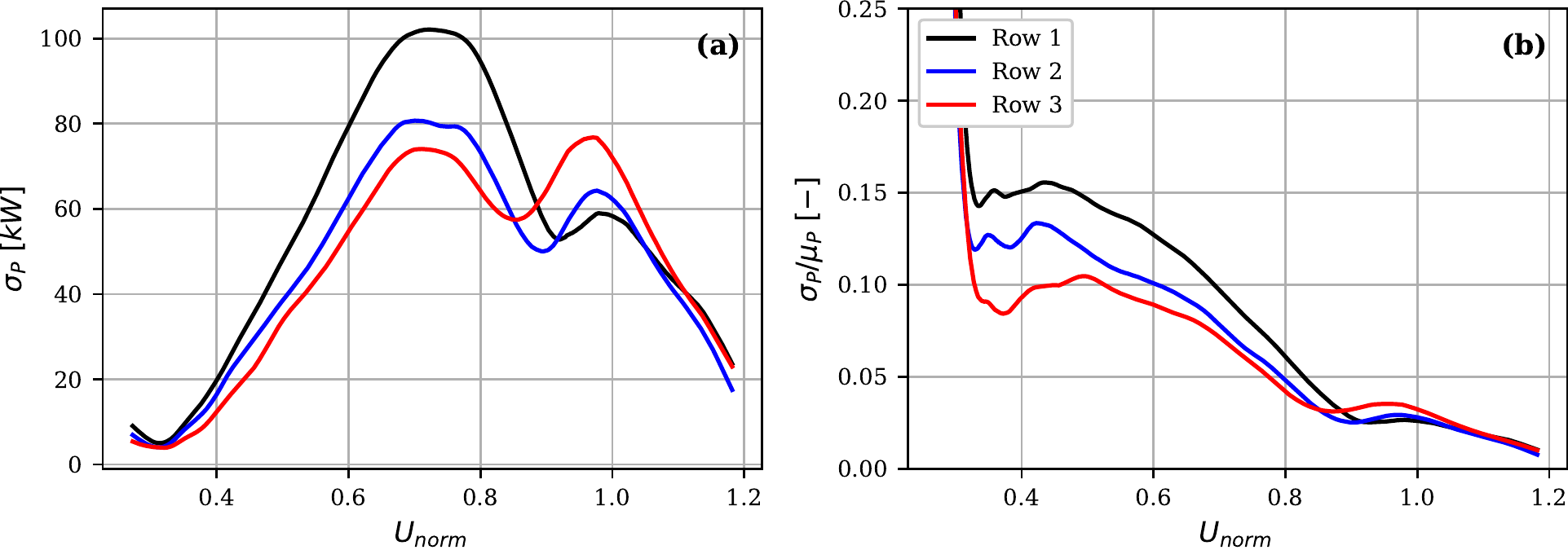}
    \caption{Sensitivity of isolated-turbine power curves aggregated by row to $TI$ estimated through the power standard deviation (a) and power standard deviation normalized by power mean (b) as a function of wind speed.}
    \label{fig:row_avg_ti_variation}
\end{figure*}

As documented in previous works \cite{El-Asha2017}, the power generated increases with $TI$ since more kinetic energy is available in the wind field. Furthermore, in region two, the curves generated by the highest and lowest $TI$ values are separated by up to nearly 400kW, further highlighting the need to accurately predict $TI$ locally at each turbine to accurately predict turbine power. Finally, the variability of the power curve with $TI$ can be seen intuitively as the vertical width of the band of generated power curves. More exactly, the standard deviation of power at a given wind speed for all the $TI$ values can be defined. This allows the sensitivity of the power curve to $TI$ variations to be defined as a function of wind speed. Performing this analysis for the isolated-turbine models aggregated on specific rows produces the result highlighted in figure \ref{fig:row_avg_ti_variation}(a). Interestingly, while the generated curves look similar, the turbine models indicate that the southernmost wind turbine row, i.e. row 1, has higher sensitivity to $TI$ than the other two rows. Normalizing the standard deviations by the mean power as in figure \ref{fig:row_avg_ti_variation}(b) does not change the results. The motivation for this feature likely is caused by the typical operation of the turbines of row 1 facing the freestream, so that they are affected by $TI$ variability associated with the daily cycle of atmospheric stability \cite{Iungo2014, El-Asha2017}. In contrast, turbines on rows 2 and 3 are significantly affected by wake interactions and, thus, are more sensitive to wake-generated turbulence rather than the incoming freestream conditions. As the isolated-turbine models do not consider waked conditions, the second and third rows are seen as less sensitive to $TI$ variations.

\section{Model Verification}\label{sec:model_verification}

Since the overall, qualitative performance of the models has been validated in the previous section, it is now important to verify that the models perform accurately in a quantitative sense. Additionally, given the availability of other studies performed on the wind farm under investigation with other modeling methods \cite{Letizia2022}, cross-comparisons between different modeling approaches are possible. 

\begin{figure*}[b!]
    \centering
    \includegraphics[width=\textwidth]{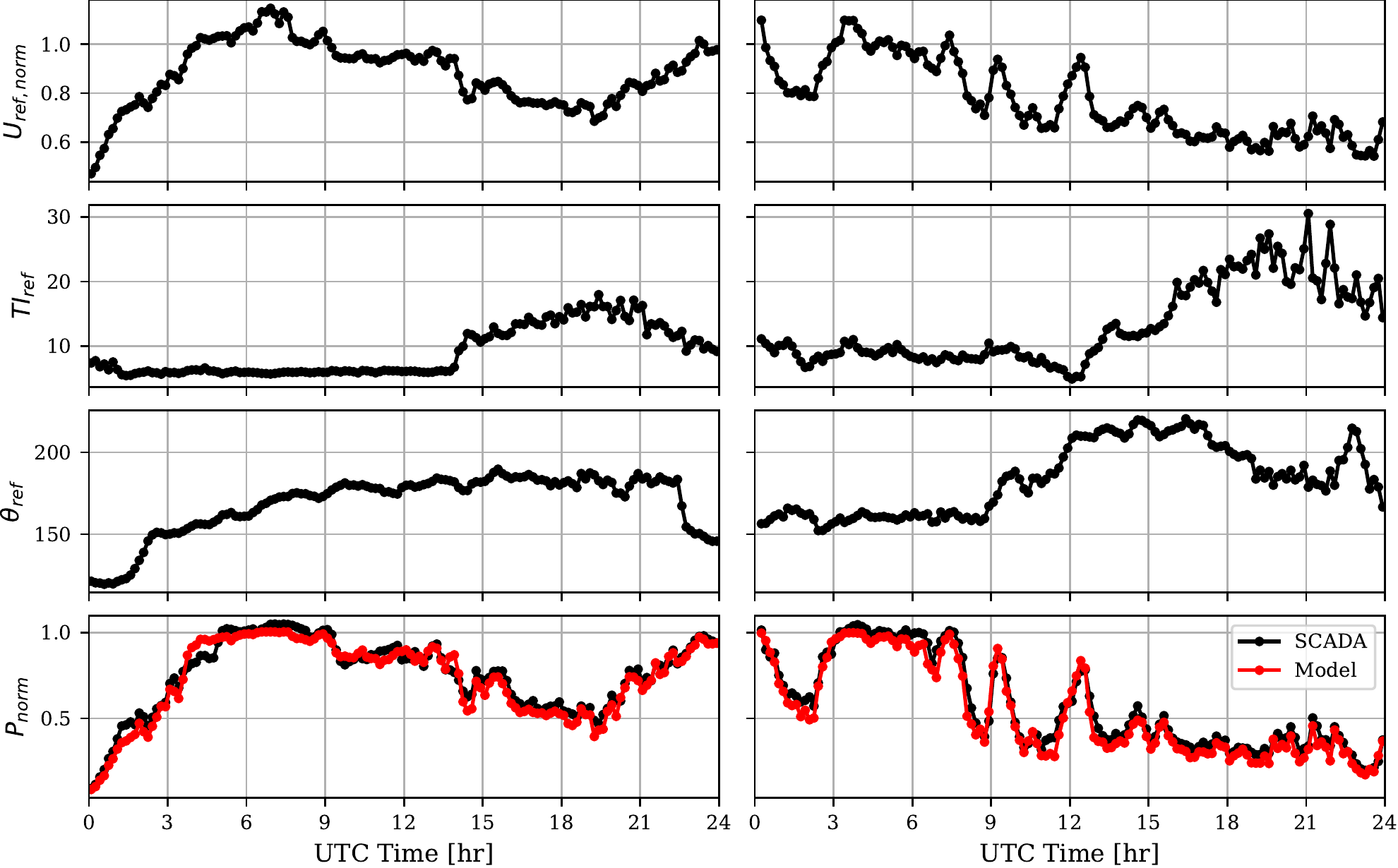}
    \caption{SCADA data from November 30\textsuperscript{th} 2016 (left column) and June 12\textsuperscript{th} (right column), 2016, with model predicted power.}
    \label{fig:timeseries_analysis}
\end{figure*}

First, the performance of the model is evaluated against the time series of SCADA data. For this analysis, the SCADA data are used to compute a time series of the total wind farm power, which is the sum of power across all the turbines at a given time. Next, the clustered-turbine model is used to predict individual turbine power capture providing as inputs the reference conditions corresponding to the time stamps used to calculate the actual total power time series. The experimental and model-predicted time series of total power capture can then be compared, as done in figure \ref{fig:timeseries_analysis} for the full days November 30\textsuperscript{th}, 2016 and June 12\textsuperscript{th}, 2016, in UTC. In both cases, the reference environmental conditions are plotted, with the reference wind speed normalized by the rated wind speed of 11 m$\cdot$s$^{-1}$, $U_{ref,~norm}$. The SCADA power reported is the total farm power normalized by the total rated power, which is the rated power of 2300 kW times 25, the total number of wind turbines, $P_{norm}$. The date of November 30\textsuperscript{th} is chosen to highlight wind farm performance under relatively steady conditions with high wind speed, low $TI_{ref}$, and southerly wind direction, $\theta_{ref}$, with the main feature being an increase in $TI$ associated with daytime, while the date of June 12\textsuperscript{th} is chosen for the increased variability in wind speed, $TI$, and direction. Overall, the model provides excellent predictions over the time series considered.

Next, the overall power and AEP predictions are compared against the true values. Summing across all time series values for actual and model-predicted total power, then calculating average yearly power production, allows the AEP to be calculated \cite{Pena2018}. Comparing AEP allows the accuracy of farm simulations to be verified at the most coarse level. For both calculations, the appropriate region removal must be applied to the actual power data points to ensure that the model is only predicting over relevant input data. Performing this analysis, the NMAE between the actual and predicted total power is 6.6\%, and the percentage difference between the actual and predicted AEP values is 3.6\%.

\begin{figure*}
    \centering
    \includegraphics[width=0.75\textwidth]{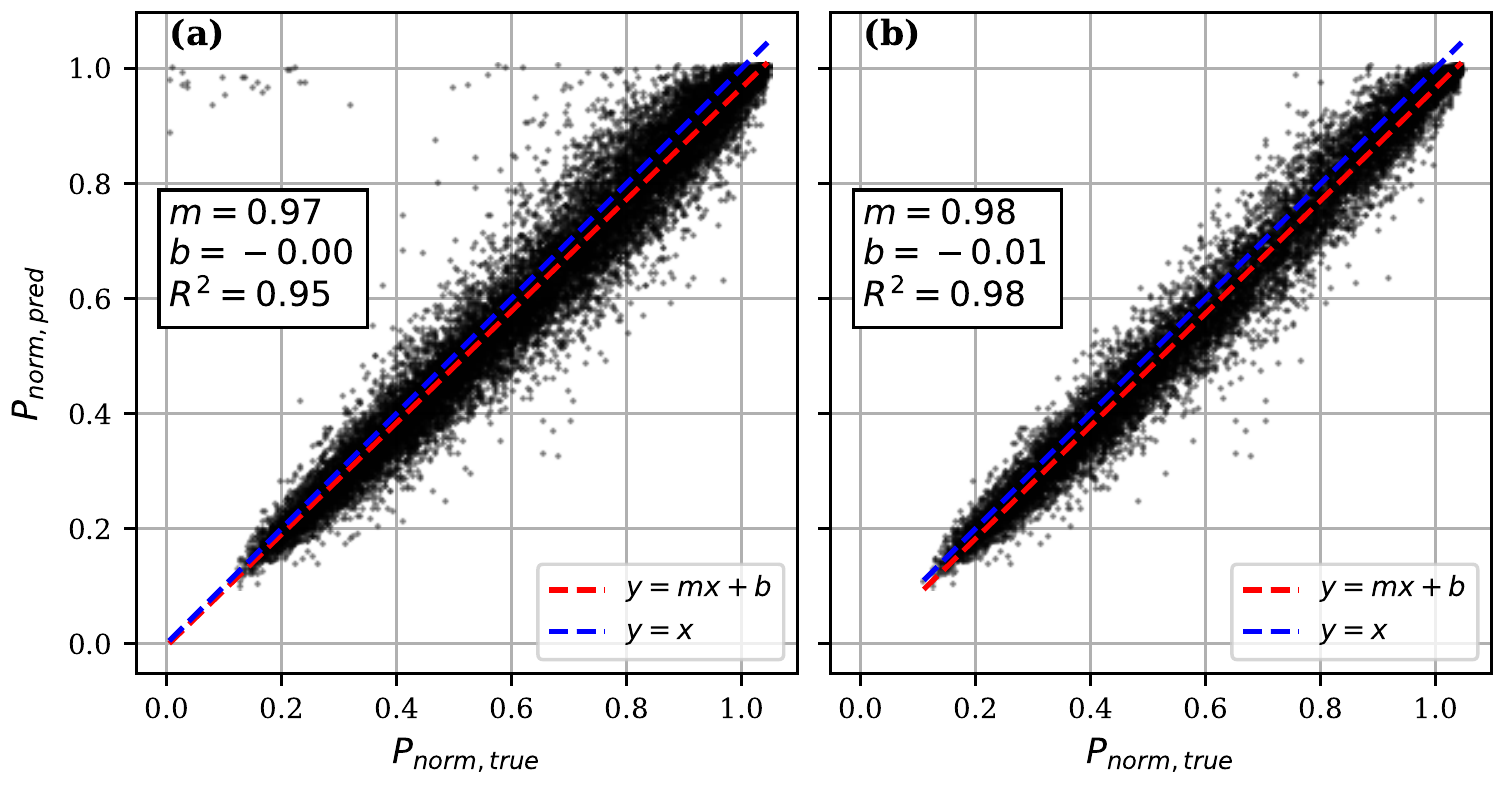}
    \caption{Regression analysis for actual and predicted total power using data with region removal applied and (a) no filter and (b) DNN filter with wakes included.}
    \label{fig:tot_pwr_regression}
\end{figure*}

To better understand the source of errors in the total power predictions, a regression analysis is performed and reported in figure \ref{fig:tot_pwr_regression}(a). While the \(R^2\) value and slope are both good, they are less accurate than similar scores for other wind farm simulation techniques, such as RANS simulations performed for the same wind farm \cite{Letizia2022}. A possible contributor to the NMAE and the lower \(R^2\) appears to be points that lie far above the \(y=x\) curve. Since these scattered points have higher predicted power than actual power, it is reasonable to guess that these may be points where one or more turbines were curtailed or were operating at lower-than-typical performance.

De-rated conditions are removed to better quantify the model accuracy, and the total power analysis is repeated for data with region removal applied as well as DNN filtration, but not wake removal. The total time series error, in this case, is 6.1\%, and the AEP error jumps to 4.2\%. From figure \ref{fig:tot_pwr_regression}(b), the cloud of points above the \(y=x\) curve - supposed to be de-rated conditions - is reduced. Noting that the slopes in figure \ref{fig:tot_pwr_regression}(b) are slightly improved over figure \ref{fig:tot_pwr_regression}(a) and that the \(R^2\) score improves, it can be concluded that this analysis gives a more accurate quantification of the performance of the model. The model itself tends to underestimate power slightly, as shown by the slope having a value slightly less than one. Thus, the outliers in figure \ref{fig:tot_pwr_regression}(a) may have improved the AEP and total power errors by slightly offsetting this underestimating trend, giving improved performance not attributable to the model. 

\begin{figure*}
    \centering
    \includegraphics[width=\textwidth]{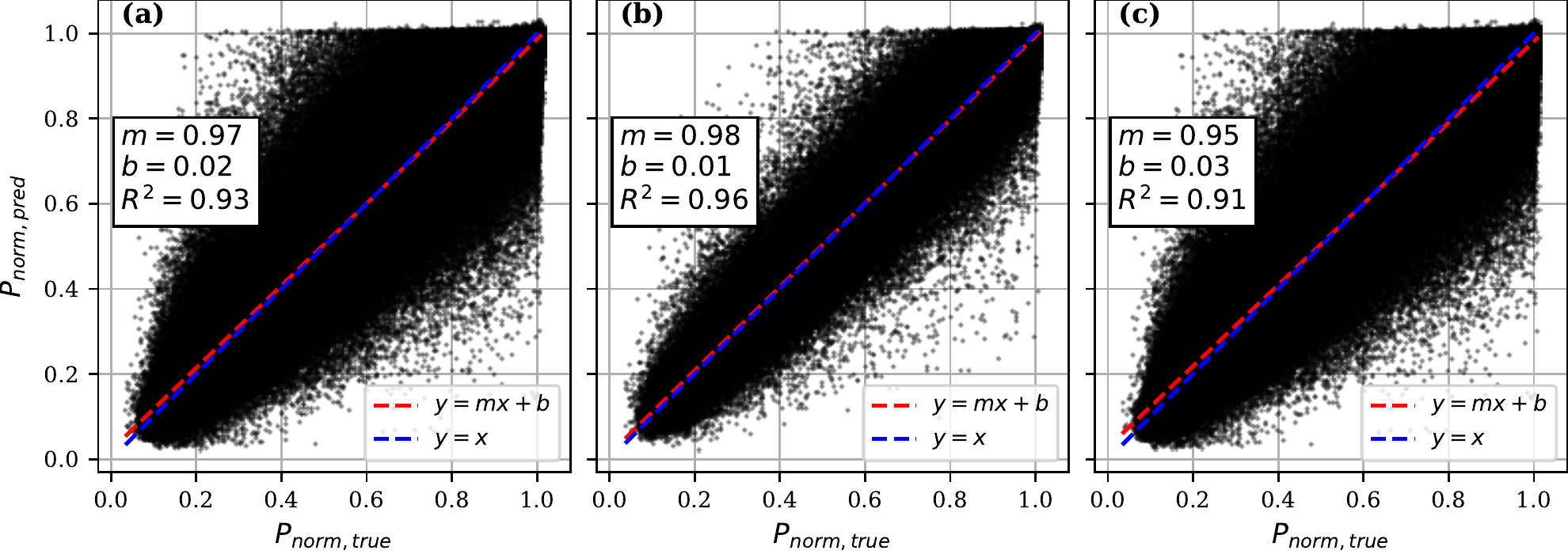}
    \caption{Regression analysis for actual and predicted power for all individual turbines combined with region removal, DNN filter applied while keeping wakes, and model refinement applied considering (a) all conditions, (b) unwaked conditions, and (c) waked conditions.}
    \label{fig:all_turbine_pwr_regression}
\end{figure*}

Next, considering the predicted power for individual turbines across the farm gives a more granular analysis. For this analysis, the real data from every turbine is filtered and the region removal procedure is applied, while still keeping wakes as before. For each resultant filtered set, the relevant reference conditions are provided as inputs to the clustered-turbine models, and the predicted power is determined. Performing this analysis, the resultant NMAE across all turbines is 9.1\%. The regression results are displayed in figure \ref{fig:all_turbine_pwr_regression}(a).

These scores highlight the complex nature of DNN predictions. The turbine power, wind speed, and $TI$ models have median NMAEs of around 3.2\%, 6.2\%, and 14.3\%, respectively, from sections \ref{sec:sub_turbine_models} and \ref{sec:sub_turbine_ws_ti_models}. The NMAE calculated here is lower than the sum of the three component models since the component predictions are not all weighted equally in the final power prediction.

To understand the impact of wakes on the model accuracy, the data considered previously are split into waked and unwaked sets, and the accuracy and regression analyses are applied again, independently, to each set. The unwaked set has an NMAE of 6.8\% across all turbines and regression results are shown in figure \ref{fig:all_turbine_pwr_regression}(b) while the waked set has an NMAE of 10.8\% and regression results are shown in figure \ref{fig:all_turbine_pwr_regression}(c).

\begin{figure*}
    \centering
    \includegraphics[width=\textwidth]{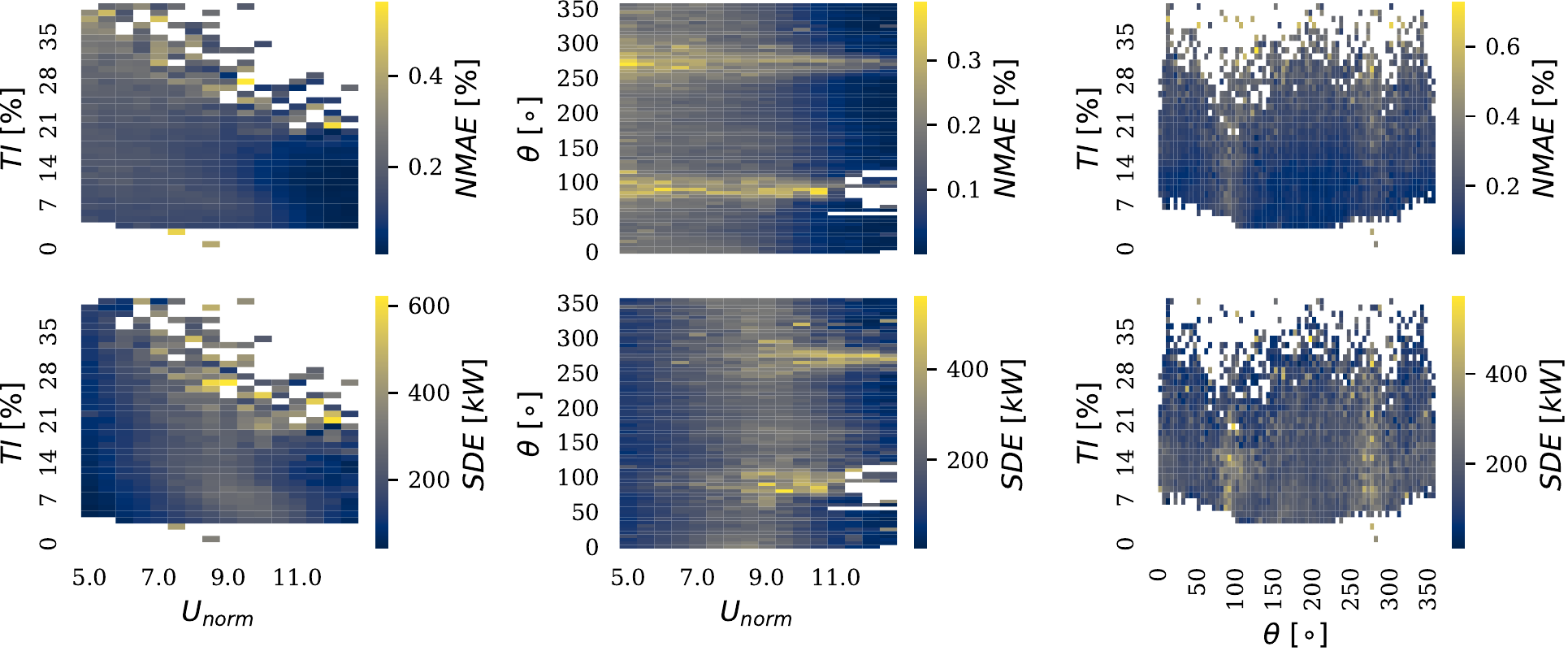}
    \caption{NMAE and the standard deviation of the error (SDE) for the clustered-turbine approach predicting power for all turbines.}
    \label{fig:all_turbine_bin_errors}
\end{figure*}

Waked samples drive prediction errors. Recalling figure \ref{fig:err_vs_wd_turb_7_8}, the wind speed model tends to overestimate the wind speed in waked sectors. While this effect is certainly strongest at low wind speeds, which have been removed from analysis during region removal, some overestimate remains for all wind speeds. Additionally, the strongest wake interactions are near-wake interactions within rows of turbines, which correspond to winds along the east-west direction. Unfortunately, data availability is lower for these wind sectors. Thus, predicting wakes has the double challenge of systematic overestimation and poor data availability. The clear impact of wakes on model performance can be seen in figure \ref{fig:all_turbine_bin_errors}. In the figure, the power is predicted for each turbine for all reference conditions from the available SCADA data. The results are then binned in three ways: wind speed and $TI$, wind speed and wind direction, and wind direction and $TI$. For each bin, the NMAE is calculated. The standard deviation in the difference between true and predicted power is also calculated and reported as the standard deviation of the error (SDE). This analysis highlights the fact that near wakes, corresponding to the east-west wind directions, tend to drive errors. These regions also tend to have the highest standard deviation in error, suggesting that the models struggle to capture these interactions the most. Despite this, the accuracy of unwaked predictions is similar to other simulation techniques applied to this site \cite{Letizia2022}.

\section{Data Requirements for Wind Farm Machine Learning Models} \label{sec:data_requirements}

Now that the performance of wind farm ML modeling has been evaluated for simulating the complex behavior of an entire wind farm, a question is posed: what are the requirements of a data set such that it can be used to train accurate wind farm ML models? For instance, given a SCADA data set, is there a metric that can indicate if it is suitable to train ML models? As another example, if an experimental campaign is being designed with ML modeling in mind, what guidelines can be implemented to ensure it would produce a data set useful for ML modeling?

To answer these questions, consider that an ML model needs sufficient data to characterize the predictable and repeatable operating conditions, otherwise, it will extrapolate from the surrounding available data. Thus, the training data must sufficiently cover the region of interest where the model with be interrogated. Furthermore, consider that ML seeks to map inputs to outputs by using features in the training data. The noisier or more variable the output is compared to the variability of the input, the more difficult the prediction becomes. In other words, variability in the output that is not correlated to variability in the inputs can increase model error. Thus, the solution should seek a metric that encompasses both the availability of the training data and the variability of the output.

The standard deviation of the power normalized by the mean of the power is selected as the indicating metric. As justification for selecting this parameter, it is noted that the NMAE is the absolute error normalized by the magnitude of power, thus using the standard deviation normalized by the magnitude of power follows a similar non-dimensionalization scheme. This metric is to be applied on a bin basis defined by the inputs to the predictive model. As noted, uncorrelated variability between the inputs and the output is theorized to lead to model error. If the bins are kept small enough that variability in the input parameters should produce theoretically negligible variability in the output parameter, then the uncorrelated variability can be isolated. 

\begin{figure}
    \centering
    \includegraphics[width=\textwidth]{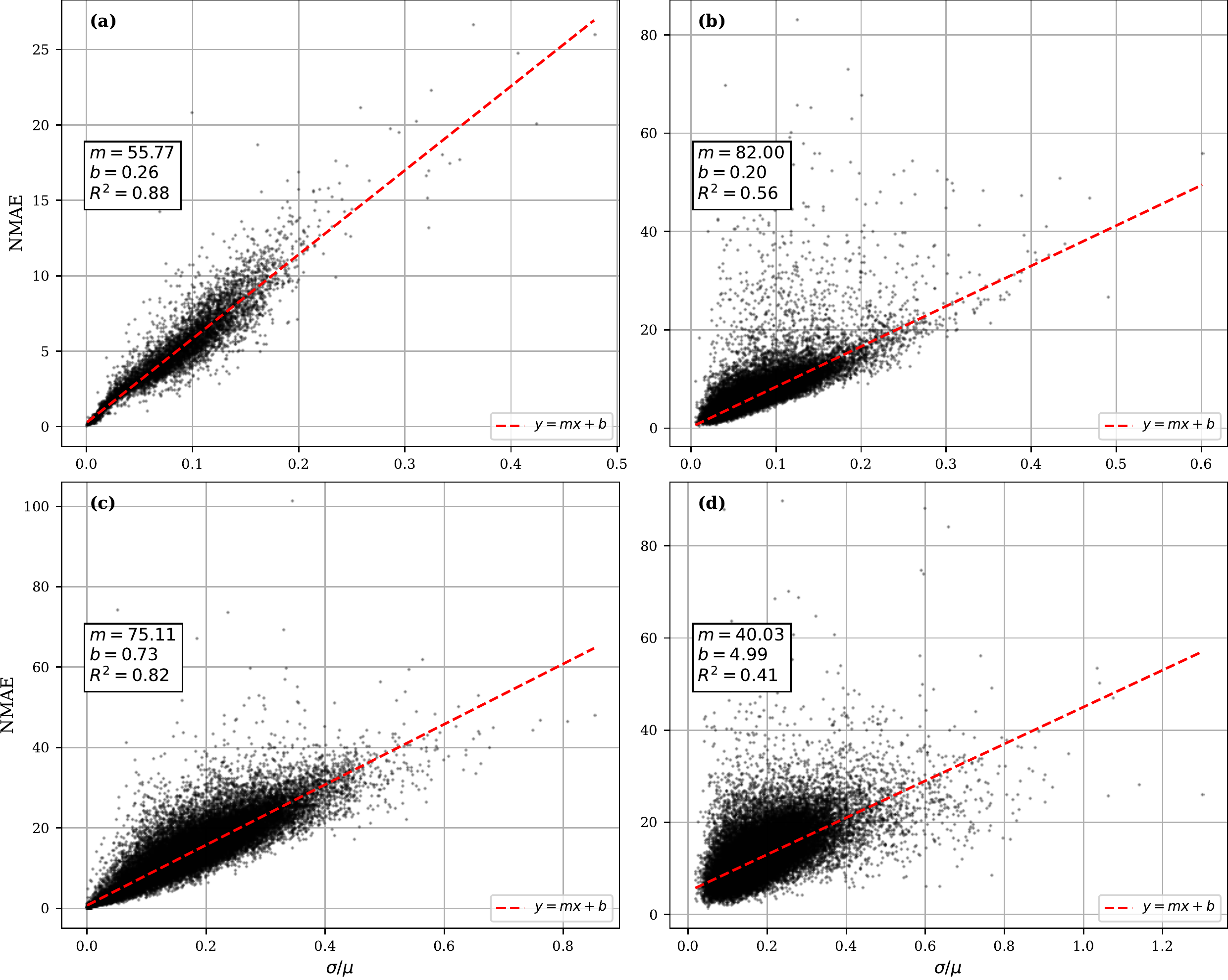}
    \caption{Bin NMAE plotted against bin normalized power standard deviation for all turbines considering turbine (a) power predicted from  SCADA data, (b) wind speed predicted from reference conditions, (c) power predicted from reference conditions, and (d) $TI$ predicted from reference conditions.}
    \label{fig:combined_NMAE_model}
\end{figure}

For this work, the inputs are reference wind speed and $TI$ and the range of interest is wind speed from 3 m$\cdot$s$^{-1}$ to 13 m$\cdot$s$^{-1}$ and $TI$ from 0\% to 35\%. Bins of interest should now be defined such that the bin is small enough so that the output is not expected to vary greatly within each bin while still making the bins as large as possible to promote statistical convergence. For instance, the IEC standard defines wind speed bins to be 0.5 m$\cdot$s$^{-1}$ wide for WTPC modeling \cite{IEC2017}. $TI$ are defined to be 2.5\% wide. For every turbine, power is predicted from the real SCADA data, thus evaluating the turbine power model only, and the NMAE and normalized power standard deviation are calculated for the bins defined above. Performing a linear regression between the two results in figure \ref{fig:combined_NMAE_model}(a), a strong linear relationship is observed with high $R^2$ suggesting that the proposed metric is indeed useful. If the NMAE of each bin in SCADA data can be accurately estimated, then the overall NMAE can be roughly estimated from equation \ref{eqn:bin_NMAE}, which defines the bin-averaged NMAE.
\begin{equation}\label{eqn:bin_NMAE}
    Bin\ Averaged\ NMAE = \frac{\Sigma_{i=1}^{N}NMAE_{i}\cdot n_i}{\Sigma_{i=1}^{N}n_i},
\end{equation}
where \(NMAE_i\) is the NMAE of the i\textsuperscript{th} bin, \(n_i\) is the number of points in the i\textsuperscript{th} bin, and \(N\) is the total number of bins. Thus, the bin-averaged NMAE can be computed using NMAE values predicted by the linear model and then compared against NMAE from the actual ML predictions. To test the accuracy of this model, the slope is 55.8, and the intercept is 0.3. Thus, the NMAE of a given bin is estimated as \(55.8\sigma/\mu + 0.3\) where $\sigma$ is the standard deviation of power in the bin and $\mu$ is the mean of power in the bin. Note that reducing the normalized standard deviation in power should theoretically reduce the NMAE, which means since the intercept of the linear regression is close to zero, the linear model recreates the expected error behavior. For the current case, the minimum, median, and maximum differences between real and bin-averaged NMAEs are reported in table \ref{tab:nmae_ws_ti_comparison}. It is noteworthy that no turbine has a bin-averaged NMAE smaller than the overall ML NMAE. Thus, if this approach is used, the bin-averaged NMAE is very likely an upper bound on model performance, and the trained model will most likely perform better than estimated.

\begin{table}
    \centering
    \begin{tabular}{|c|c|c|c|}
        \hline
         & \textbf{Min} & \textbf{Median} & \textbf{Max} \\ \hline
         \textbf{Power from SCADA NMAE} & 0.99\% & 1.22\% & 1.63\% \\ \hline
         \textbf{WS NMAE} & -0.34\% & 0.05\% & 0.36\% \\ \hline
         \textbf{$\mathbf{TI}$ NMAE} & 3.88\% & 3.24\% & 3.93\% \\ \hline
         \textbf{Power from reference NMAE} & 3.64\% & 4.15\% & 5.33\% \\ \hline
    \end{tabular}
    \caption{Statistics of overall NMAE minus predicted NMAE for wind speed and $TI$ for all turbines.}
    \label{tab:nmae_ws_ti_comparison}
\end{table}

The above approach is now applied to the models for turbine wind speed and $TI$. The previous wind speed limits are removed for these models, so the wind speed range is from 0 m$\cdot$s$^{-1}$ up to 25 m$\cdot$s$^{-1}$ with 0.5 m$\cdot$s$^{-1}$ wide bins. The 1\textsuperscript{st} and 99\textsuperscript{th} percentiles in reference $TI$ are 4.7\% and 49\% respectively, so the $TI$ limits are set to 5\% and 47.5\% with 2.5\% wide bins. Wind direction is from 0\(^{\circ}\) to 360\(^{\circ}\) with 5\(^{\circ}\) wide bins. The resultant linear regression between bin NMAE and bin normalized standard deviation in turbine wind speed and $TI$ is shown in figure \ref{fig:combined_NMAE_model}(b) and (d). The linear relationship observed between bin NMAE and bin normalized standard deviation for power predictions is not significantly replicated in the wind speed and $TI$ predictions. This is likely because wind speed and $TI$ are more difficult overall to predict accurately than power. Nevertheless, using the linear fits, a bin-averaged NMAE is computed for each turbine's wind speed and $TI$ predictions and compared with the overall NMAE. Statistics on the real NMAE minus the bin-averaged NMAE are shown in table \ref{tab:nmae_ws_ti_comparison}.

Though wind speed and $TI$ linear fits did not appear very accurate, the NMAE is still predicted with remarkable accuracy. Thus, the overall power when predicted from reference conditions using combined global and local models should also be accurately predicted using the NMAE modeling approach. Using the same binning approach as for the global wind speed and $TI$ models, with the exception that the minimum wind speed is set to 5 m$\cdot$s$^{-1}$, the bin NMAE and normalized power standard deviation are calculated for all bins over all turbines and combined in figure \ref{fig:combined_NMAE_model}(c) to generate the linear fit. Though the fit is not as good as the local power model fit, it is still far better than the global wind speed and $TI$ fits and suggests that the linear NMAE model may be appropriate in this case. Once again, the intercept is close to zero, indicating the model is recreating expected NMAE behavior. Calculating the bin-averaged NMAE for all turbines as well as the overall NMAE combining all the considered bins for all turbines, the minimum, median, and maximum predicted minus overall NMAE values are reported in table \ref{tab:nmae_ws_ti_comparison}. Similarly to the power models predicting on real SCADA data, the predicted NMAE is always greater than the overall NMAE. Thus, if the NMAE linear model is used to predict model performance from a given data set, it is very likely that the real model should outperform the predictions. Though the generality of this approach should be tested on other turbines and other wind farms, these results show promise that the accuracy of power predictions could be estimated with an error of $5\%$ or better from the statistics of the training data set alone.

\section{Computational Cost} \label{sec:computational_cost}
Though the ML models defined thus far have been shown to be accurate in recreating farm performance, they will not be broadly useful unless they have sufficiently low computational costs. Thus, the computational cost of the models is defined and compared against a RANS solver previously applied to the site under consideration, and being a mid-fidelity model can be considered as a good trade-off between model accuracy and required computational costs \cite{Letizia2022}. 

To calculate computational time, a typical consumer desktop computer is used with an Intel Core i7-9700K CPU and an NVidia Quadro RTX 4000 GPU. First, the ML computational cost is established by training the individual-turbine wind speed and $TI$ models needed for the clustered-turbine approach. Training individual turbine models takes, on overage, 15 minutes per turbine. Thus, for 25 turbines, the average time to train the models is 6.25 hours. Similarly, wind speed and $TI$ models both take 23 minutes on average to train the models for a single turbine, meaning approximately 9.5 hours for all the wind speed and $TI$ models for the farm each. In total, all the models take 25.25 hours to train.

Conversely, to establish the computational cost for the RANS solver, 20 cases are simulated with random environmental conditions selected with wind speed between 3 m$\cdot$s$^{-1}$ and 13 m$\cdot$s$^{-1}$, $TI$ between 2.5\% and 35\%, and wind direction between 0$^{\circ}$ and 360$^{\circ}$. The average time for a single environmental case from this analysis is 30 seconds. Note that the RANS computational cost is at best a rough estimate without simulating the entire farm for all the environmental conditions of interest because some environmental conditions converge more quickly than others due to different levels of wake interactions.

While the computational cost for the RANS solver to simulate a single environmental case is extremely low compared to the cost to train all the requisite ML models, the advantage that the ML models hold is that once trained, they can be used to simulate any environmental condition for the trivially low computational cost of mere seconds on a modern laptop computer. Thus, the farm can be easily simulated for any desired condition, or all conditions, with almost no appreciable computational cost. With the RANS solver, however, a nearly identical computational cost must be paid every time a new simulation is needed. Thus, simulating the entire farm for varying environmental conditions, the cost can quickly add up. 
For example, suppose that the farm is to be investigated varying wind speed, direction, and $TI$ by choosing a modest grid using the IEC guidelines for power curves to determine the minimum steps in wind speed and wind direction. Let the wind speed vary between 3 m$\cdot$s$^{-1}$ and 13 m$\cdot$s$^{-1}$ in steps of 0.5 m$\cdot$s$^{-1}$, $TI$ between 2.5\% and 35\% in five even steps, and wind direction between 0$^{\circ}$ and 360$^{\circ}$ in steps of 5$^{\circ}$ with the knowledge that IEC recommends wake sectors to be kept 5$^{\circ}$ wide but that wake effects can also occur on a smaller scale. This results in 7560 combinations to be simulated. Assuming each case requires 30 seconds for the RANS solver to solve, this adds up to a total estimated computational cost of 63 hours. The ML approach, however, once trained, can solve these cases in mere minutes. Thus, to simulate the wind farm using the RANS solver while maintaining a similar level of detail to the solutions available via the ML approach requires more than twice the computational time of the ML approach.

Another drawback of the RANS solver is that the computational cost may increase drastically with increasing farm complexity. Since it is a physics-based solver, more challenging farm layouts, or layouts with more turbines, can greatly increase the computational cost necessary, or the likelihood that individual cases fail to converge. Conversely, the ML models, being entirely data-driven, do not scale in computational cost with the complexity of the farm but rather with the size of the data sets used to train the models. Assuming that the data sets are the same size from a small farm to a large one, then the only increase in the ML computational cost would be in training additional models for the additional turbines, and thus the computational cost will only grow linearly with the number of turbines if the data set size is held constant.

In conclusion, if only a small number of environmental condition cases are needed, or the wind farm being considered is relatively simple and small in layout, then the RANS solver will likely have a lower computation cost. If, however, there is a need to simulate the wind farm quickly for any arbitrary environmental condition or a sufficiently large distribution of conditions, such as for the optimization of the wind farm layout, or if the farm is large or complex, then the ML models will likely have a lower computational cost in the end.

\section{Concluding Remarks} \label{sec:conclusion}

Modeling wind turbines, wind farms, and wake interactions is a challenging task with several existing solutions. Most approaches using simulations have either a high accuracy at the cost of high computational cost or a low computational cost with reduced accuracy. Additionally, models providing accurate predictions of the turbulence intensity ($TI$) and its variability within a farm, such as due to wake-generated turbulence, in a practical amount of time are essentially lacking. This work instead uses machine learning (ML), which can combine lower computational cost, good accuracy, and accurate $TI$ predictions.

This paper has proposed multiple applications of ML for modeling wind turbine and wind farm operations in terms of power capture, wind speed, and $TI$ at each turbine location. Using SCADA data from a wind farm in the Panhandle of Texas, deep neural network models have been used to filter the experimental data set from outliers and off-design operations as well as select optimal inputs for further simulation. Reference conditions have been defined for the wind farm and clustered-turbine models have been used to predict local wind speed and $TI$ at individual turbine locations from the reference inputs of wind speed, direction, and $TI$. Turbine models have been used to predict turbine power, generating power curves that capture individual turbine performance and its sensitivity to $TI$. Used in combination, these models can simulate complex wake behavior such as variations in near and far wake strength, wake dependence on $TI$, and speedups with corresponding $TI$ damping.

When these models are validated over historical SCADA data, the predicted farm power tracks the real farm power almost identically. The total farm power for all data points and the annual energy production are both predicted accurately, with errors of 6.1\% and 4.2\%, respectively. At the individual turbine level, waked conditions - with a normalized mean absolute error (NMAE) score of 10.8\% across all turbines - are still slightly more difficult to predict than unwaked conditions - with an NMAE score of 6.8\% across all turbines, but overall accuracy remains high at 9.1\% NMAE. Careful removal of de-rated conditions in the SCADA data confirms the significance of these results by isolating the performance of the model under intended conditions.

Additionally, it has been considered what characteristics of a data set can be used to indicate suitability for training wind farm ML models. Using the statistics of the output parameter in bins defined by the input parameters, the NMAE of the ML predictions can be accurately estimated. Thus, data sets may be evaluated or experiments designed to produce good data sets for ML by utilizing these NMAE predictions without the need to train and evaluate ML models on the data sets themselves.

While there is an upfront computational cost to train ML models for use in simulation, the advantage is all environmental conditions are baked into the model during training. After training, the entire farm can be simulated under any given environmental condition in seconds on consumer hardware. Furthermore, there is no need to retrain for new environmental conditions, so any desired condition can be simulated very quickly compared to other methods that require a new simulation for every new environmental condition. Because of this, the computational cost of the ML approaches was found to be as much as half that of a sample mid-fidelity Reynolds Averaged Navier-Stokes solver. Finally, using the clustered-turbine approach, $TI$ has been predicted at the turbine level, a difficult feature to accomplish with current models in a computationally cost-effective manner.

Though the current modeling technique limits the models to the single wind farm on which they were trained, the promise of using ML modeling is clear. The next steps for this research will encompass developing wind farm ML models that can be generalized to wind turbines and wind farms other than those considered in the training of the models. Such models would enable powerful, data-driven farm design and optimization and would be a great benefit to the wind energy sector.

\section*{Acknowledgements}
This material is based upon work supported by the National Science Foundation (NSF), CBET Fluid Dynamics, CAREER program, Award No. 2046160, Program Manager Ron Joslin.

\section*{Data Availability Statement}
The data that support the findings of this study are protected by a non-disclosure agreement.

\end{document}